\def\bea{\begin{eqnarray}}
\def\eea{\end{eqnarray}}
\def\refeq#1{(\ref{#1})}
\def\be{\begin{equation}}
\def\ee{\end{equation}}
\def\e{{\rm e}}

\def\squareforqed{\hbox{\rlap{$\sqcap$}$\sqcup$}}
\def\qed{\ifmmode\squareforqed\else{\unskip\nobreak\hfil
\penalty50\hskip1em\null\nobreak\hfil\squareforqed
\parfillskip=0pt\finalhyphendemerits=0\endgraf}\fi}

\def\Im{{\rm Im\,}}
\def\Re{{\rm Re\,}}

\def\<{\langle}
\def\>{\rangle}

\documentstyle[prb,twocolumn,aps,rotate,epsf,floats]{revtex}

\newcommand{\tw}{\tilde{\omega}}

\def\rbx#1{\raisebox{-0.3ex}{$\scriptstyle #1$}}

\begin{document}
\setlength{\unitlength}{1cm}
\renewcommand{\arraystretch}{1.4}

\title{Line shapes of dynamical correlation functions in Heisenberg
       chains}  

\author{Ralph Werner}
\address{Physics Department, Brookhaven National
Laboratory, Upton, NY 11973-5000, USA} 

\author{Andreas Kl{\"u}mper}
\address{Physics Department, University of Dortmund, D-44221
Dortmund, Germany}

\date{\today}

\maketitle

\centerline{Preprint. Typeset using REV\TeX}

\begin{abstract} 
We calculate line shapes of correlation functions by use of complete
diagonalization data of finite chains and analytical implications from
conformal field theory, density of states, and Bethe ansatz.
The numerical data have different finite size accuracy 
in case of the imaginary and real parts in the frequency and
time representations of spin-correlation functions, respectively.
The low temperature, conformally invariant regime crosses over at
$T^*\approx 0.7J$ to a diffusive regime that in turn connects
continuously to the high temperature, interacting fermion regime. The 
first moment sum rule is determined.
\end{abstract}

%%%%%%%%%%%%%%%%%%%%%%%%%%%%%%%%%%%%%%%%%%%%%%%%%%%%%%%%%%%%%%%%%%%%%

\section{Introduction}

Dynamical correlations characterize the spectral properties of
physical systems. They are accessible by a multitude of experimental
setups. The access to dynamical correlation functions for physically
relevant systems is usually difficult even in exactly solvable
models.\cite{KM97,KHM98} Dynamical spin-correlation functions in
Heisenberg chains have been widely studied
numerically\cite{SSS97b,FLS97,YS97,FL98} as well as
analytically.\cite{MTBB81,Schu86,Tsve95} The comparison of numerical
and approximate analytical results for the purpose of accuracy control
has been used in various previous
approaches.\cite{EN88,CPK+95,YMV96,KMB+97,SSS97a,KHM00,KM00}  

Usually the focus lies on the imaginary part of the correlation
functions. The real and the imaginary parts can be Kramers-Kronig
transformed into each other and thus hold the same information. This
is also true for the Fourier transform. The information that can be
extracted from finite systems accessible by exact diagonalization (ED)
concerning the thermodynamic limit is limited. The accuracy of the
results is different for different representations. In the case of
finite systems it proves thus useful to actually calculate all three
representations to extrapolate to the thermodynamic
limit.\cite{Wern01a}   

The dynamical correlation functions become system size independent for
high excitation energies\cite{Wern01a} or, equivalently, on short time
scales.\cite{FLS97} While finite systems thus allow for the determination
of correlation functions in the thermodynamic limit at high
frequencies or on short time scales, field theoretical results
describe their asymptotic behavior on long time scales or for small
frequencies.\cite{Schu86,Tsve95} The perspective of this paper is to
combine the strongholds of both methods.

The system to be discussed here is the one-dimensional antiferromagnetic
Heisenberg model
\begin{eqnarray}\label{H}
H&=&\sum_{l}\left(J\,S^x_{l} S^x_{l+1} + J\,S^y_{l} S^y_{l+1} + 
	             J_z\,S^z_{l} S^z_{l+1}\right)
\nonumber\\&&
    +\ J_2\sum_{l}\left(S^x_{l} S^x_{l+2} + S^y_{l} S^y_{l+2} + 
	             S^z_{l} S^z_{l+2}\right)
\end{eqnarray}
with the superexchange integrals $J$ and $J_2$ between nearest-neighbor
(NN) and next-nearest-neighbor (NNN) magnetic ions, respectively,
$z$-axis anisotropy $J_z$ and spin-1/2 operator components
$S^\nu_{l}$ with $\nu=x,y,z$ at site $l$. Energies will be given in
units of the in plane exchange, i.e., $J\equiv 1$. This Hamiltonian is
relevant for the description of the magnetic systems in many
quasi-one-dimensional materials as Sr$_2$CuO$_3$,\cite{ACH+95,MEU96} 
$\mathrm{Cs_2CuCl_4}$,\cite{CTC+96} KCuF$_3$,\cite{TCNT95} or 
CuGeO$_3$.\cite{Wern99}  

We focus on the spin-correlation function
\begin{equation}\label{chi}
\chi(q,i\omega_n) = 
\frac{1}{L}\int_0^\beta d\tau
     \ {\rm e}^{i\omega_n \tau}\
     \left\langle S^z_{q}(\tau) S^z_{-q}(0) \right\rangle\
\end{equation}
with Matsubara frequencies $\omega_n=2\pi n/\beta$, inverse
temperature $\beta=1/T$, ($k_{\rm B}\equiv 1$) Fourier transformed spin
operators in interaction representation $S^z_{q}(\tau)={\rm
e}^{-H\tau} \sum_{l}{\rm e}^{-iql} S^z_l\ {\rm e}^{H\tau}$, and number
of sites $L$. In its analytically continued form, where $i\omega_n \to
\omega + i\epsilon$ with $\epsilon\to 0$, it determines the structure
factor  
\begin{equation}\label{structure}
S (q,\omega) = \frac{1}{\pi}\ \frac{{\rm Im} \chi(q,\omega)}
	{1-{\rm e}^{-\beta\omega}}
\end{equation}
relevant for neutron scattering experiments.

\subsection{Numerical methods}

For finite systems the correlation function can be calculated 
through the diagonalization of the spin Hamiltonian in the spectral
representation since eigenfunctions $|n\rangle$ and eigenvalues $E_n$
are known. All numerical results in this paper are obtained using
periodic boundary conditions. Defining the matrix elements
\begin{equation}
V_{nm}=\left\langle n\left| S^z_{q} \right| m \right\rangle
\end{equation}
and the Boltzmann factor
\begin{equation}
f_{nm}(\beta)=\frac{1}{Z}({\rm e}^{-\beta E_n}-{\rm e}^{-\beta E_m}),
\end{equation}
where $Z={\rm Tr}\;{\rm e}^{-\beta H}$ is the partition function, one
can write 
\begin{eqnarray}
\label{Rechispectral}
{\rm Re}\,\chi(q,\omega) &=& -\lim_{\epsilon\to 0}\sum_{m,n} 
\frac{f_{nm}(\beta)\left|V_{nm}\right|^2\! (\omega+E_n-E_m)}
     {(\omega+E_n-E_m)^2+\epsilon^2},
\nonumber\\[-2ex]
\\[1ex]
\label{Imchispectral}
{\rm Im}\,\chi(q,\omega) &=& \pi \sum_{m,n} 
f_{nm}(\beta)\left|V_{nm}\right|^2\ 
     \delta(\omega+E_n-E_m).
\nonumber\\[-2ex]
\end{eqnarray}

The corresponding real-time retarded spin-correlation function is
obtained via a Fourier transformation as 
\begin{eqnarray}\label{Realtimefinite}
\chi (q,t) &=& 
        \frac{1}{2\pi}\int_{-\infty}^\infty d\omega\ 
        {\rm e}^{-i\omega t}\ \chi_{\rbx{}}(q,\omega)\
\nonumber\\&=& - i\, \theta(t)
\sum_{m,n} f_{nm}(\beta)\ \left|V_{nm}\right|^2\ 
                 {\rm e}^{i(E_n-E_m)t},
\end{eqnarray}
where $\theta(t)$ is the Heaviside function. 

To determine the correlation functions in frequency space we use the
same methods as described in Ref.\ \onlinecite{Wern01a} which we briefly
summarize. At low temperatures small systems exhibit a small number of
dominant spectral lines at frequencies $\tw_j$ which usually can be
attributed to specific excitations.\cite{KHM00,KM00} The imaginary
part of the correlation function is determined most accurately by 
``binning'' the  data as 
\begin{eqnarray}\label{Imchibins}
{\rm Im}\,\chi(q_z,\tw_j^{\rm inf}<\omega<\tw_j^{\rm sup}) &=& 
\nonumber\\&&\hspace{-27ex}
\pi\!\! \sum_{m,n}\! \frac{f_{nm}(\beta)\left|V_{nm}\right|^2 
                 [\theta(\omega_{nm}-\tw_j^{\rm inf}) -
                  \theta(\omega_{nm}-\tw_j^{\rm sup})]}
{\tw_j^{\rm sup}-\tw_j^{\rm inf}}.
\nonumber\\[-2ex]
\end{eqnarray}
For small systems at low temperatures the appropriate choice is such
that the interval boundaries lie in the middle between the dominant
spectral lines:   
\begin{equation}\label{defomlim}
\tw_j^{\rm sup} = \tw_{j+1}^{\rm inf} = (\tw_{j}+\tw_{j+1})/2\,.
\end{equation}

If only the ``dominant'' spectral lines are present and if those lines
form a well defined continuum in the thermodynamic limit, i.e., for
$L\to\infty$, Karbach, M{\"uller}, and coworkers have
shown\cite{KMB+97,KHM00,KM00} that Eq.\ (\ref{Imchispectral}) can be
used, appropriately scaled to the thermodynamic limit, by introducing
a density of states with respect to appropriate quantum numbers
derived from Bethe ansatz. This leads to the following representation
of the imaginary part of the correlation  
function:\cite{Wern01a} 
\begin{equation}\label{Imchidens}
{\rm Im}\,\chi(q_z,\tw_j) =
\sum_{m,n}^{\tw_j=E_n-E_m} 
            \frac{2\pi\ f_{nm}(\beta)\left|V_{nm}\right|^2}
                 {\tw_{j+1}-\tw_{j-1}}.                 
\end{equation}
The sum covers only values of $n$ and $m$ such that $\tw_j=E_n-E_m$.
In Heisenberg chains this representation is only applicable at $T=0$. 

It can be shown that Eq.\ (\ref{Rechispectral}) gives very accurate
results for the real part of the correlation function if it is
determined at the dominant spectral lines $\tw_j$.\cite{Wern01a}
\begin{eqnarray}
\label{Rechiregular}
{\rm Re}\,\chi(q_z,\tw_j) &=& 
\nonumber\\&&\hspace{-12ex}
-\sum_{m,n} \frac{f_{nm}(\beta)\left|V_{nm}\right|^2}
     {(\tw_j+E_n-E_m)}\ \theta(|E_n-E_m-\tw_j|-\Delta\omega)\!\!
\end{eqnarray}
The regularization parameter $\Delta\omega$ can be set to zero if only
excitations at $\tw_j$ are present (define $\theta(0)=0$). For
Heisenberg chains at intermediate temperatures and frequencies a 
choice of $\Delta\omega=0.1J$ yields reliable results. For higher 
frequencies the results for the real part of the correlation functions
are free of finite size effects.   

%%%%%%%%%%%%%%%%%%%%%%%%%%%%%%%%%%%%%%%%%%%%%%%%%%%%%%%%%%%%%%%%%%%

\subsection{Field theoretical preliminaries and
transformations}\label{sectionCFT}

The correlations described by $\chi(q,\omega)$ Eq.\ (\ref{chi}) are
dominant at $q=\pi$ reflecting the antiferromagnetic instability of
the system. We will thus focus on this wave vector. For $q\sim\pi$
and $J_2=0$ the spin-correlation function has been studied in detail
with bosonization techniques by Schulz\cite{Schu86} and has later been
improved including logarithmic corrections.\cite{SSS97a} The result
of conformal field theory for any two-point function with scaling dimension
$x$ in Euclidean space $(r,\tau)$ at low temperature $T$ is\cite{Tsve95}
\be\label{realspaceCFT}
\chi_{\rbx{\rm CFT}}(r,\tau)=\chi_{\rbx{0}}\left[
\frac{\frac{\pi T}v}{\sinh\pi T\left(\frac rv+i \tau\right)}
\frac{\frac{\pi T}v}{\sinh\pi T\left(\frac rv-i \tau\right)}
\right]^x,
\ee
where $v$ denotes the velocity of the low lying spin excitations, and
$\chi_{\rbx{0}}$ is some constant. The spin wave velocity for
frustrated Heisenberg chains has been determined numerically as $v=0.5
\pi (1-1.12J_2)$ for $J_2<0.2411$.\cite{FG97} The Fourier representation
in momentum $q$ and frequency $\omega$ space with $\Im\omega>0$ is
\bea\label{ChiCFT}
\chi_{\rbx{\rm CFT}}(q,\omega)&=&
\sin(\pi x)\ v^{1-2x}\ \chi_{\rbx{0}} \left(\pi T\right)^{2x-2} 
\\[0.5ex]\nonumber
&&\hspace{-6ex} F_x\left(\frac{\omega-v(q-\pi)}{2\pi T}\right)
F_x\left(\frac{\omega+v(q-\pi)}{2\pi T}\right)\!\!\!
\eea
with
\bea\label{function}
F_x(k)&=&\int_0^\infty d\lambda\frac{\e^{i\lambda
k}}{(\sinh\lambda)^{x}}
\nonumber\\
&=&2^{x-1}\Gamma(1-x)\frac{\Gamma(x/2-ik/2)}{\Gamma(1-x/2-ik/2)}.
\eea
The value of the scaling dimension $x$ depends on the strength of the
interaction or the anisotropy in case of a spin chain. For the $XY$
model we have $x=1$, and for the isotropic Heisenberg chain
$x=1/2$. In the lower half plane $\chi_{\rbx{\rm CFT}}(q,\omega)$ is
given by Eq.\ \refeq{ChiCFT} with $\omega$ replaced by $-\omega$.

From this representation we learn that the function on the right hand
side of Eq.\ \refeq{ChiCFT} is analytic in a strip around the real
axis with $|\Im \omega| < x 2\pi T$ as long as $T>0$; for $T=0$ we
have  
\be\label{ChiTzero}
\Im\chi_{\rbx{\rm CFT}}(q,\omega) \simeq \cases{
0, & \hspace{-20ex} for $\omega<v|q-\pi|$ \cr
\left[\omega^2 - v^2(q-\pi)^2\right]^{x-1}, & else.}
\ee 
These analytical properties are shared by the structure factor $S_{\rm
CFT}(q,\omega)$, which is related to $\Im \chi_{\rbx{\rm
CFT}}(q,\omega)$ via Eq.~(\ref{structure}), i.e., it is analytic in
$|\Im \omega| < x 2\pi T$ for $T>0$, and $S_{\rm
CFT}(\pi,\omega)\simeq \omega^{2x-2}$ for $T=0$.  

Performing the Fourier transform to real time we see that both
$\chi_{\rbx{\rm CFT}}(q,t)$ and $S_{\rm CFT}(q,t)$ decay
exponentially at finite temperatures and algebraically for $T=0$ and
$q=\pi$,
\be\label{timelimitCFT}
\chi_{\rbx{\rm CFT}}(\pi,t) \simeq \cases{
   \exp(-x 2\pi T t),& for $T>0$ \cr
   t^{1-2x},& for $T=0$.}
\ee
For momenta $q\not=\pi$ the function $\chi_{\rbx{\rm CFT}}(q,t)$
decays exponentially with time $t$ for any $T>0$ as well as $T=0$.

There are additional contributions to $\chi(q,\omega)$ and
$S(q,\omega)$ on the lattice that are singular at finite values of
$\omega$ even for $T>0$. These contributions have their origin in the
existence of the lattice which leads to finite energy bands with upper
band edge singularities. There are no universal predictions like for
the lower band edge governed by conformal field theory and described
above. An exception, of course, is the $XY$ spin model which can be
mapped to free fermions.

As we discuss in Sec.\ \ref{sectionXYmodel} the case of the $XY$ model
suggests to assume that $\chi(q,\omega)$ is singular at a frequency
$\Lambda$ where the imaginary part diverges like  
\be
\Im\chi(q,\omega\pm i\epsilon)=\cases{
\pm(\Lambda-\omega)^{\alpha},& for $\omega<\Lambda$,\cr
0,& for $\omega>\Lambda$.\cr}
\ee
The upper (lower) sign yields the retarded (advanced) correlation
function. If not stated explicitly we discuss the retarded functions.
The Kramers-Kronig transform yields the singularity of the real part
\be\label{KKcutoff}
\Re\chi(q,\omega)=\cases{
\cot\pi\alpha(\Lambda-\omega)^{\alpha},& for $\omega<\Lambda$,\cr
\frac 1 {\sin\pi\alpha}(\omega-\Lambda)^{\alpha},& for $\omega>\Lambda$.\cr}
\ee
In the neighborhood of $\alpha=0$ we have a logarithmic
singularity\cite{Wern01a}
\be\label{Reln}
\Re\chi(q,\omega)=\frac 1 {\pi}\ln|\Lambda-\omega|.
\ee
Regarding the time dependence we note that both functions
$\chi (q,t)$ and $S (q,t)$ are dominated by the
singularity at $\Lambda$ and show long time asymptotics 
\be\label{FTcutoff}
\chi (q,t) \simeq t^{-(1+\alpha)}\exp(-i\Lambda t).
\ee 

Since the operator $S^z$ is self adjoint $\Im\chi(\pi,\omega)$ is odd
in $\omega$. In general we thus set $\Im \chi (q,\omega) \sim {\rm
sign}(\omega)\,(\Lambda^2 - \omega^2)^\alpha$. The Fourier transform
FT$[\chi(\pi,\omega)]$ is consequently identical to twice the sine
transform of $\Im\chi(\pi,\omega)$ and $\chi (\pi,t)$ is
real. For the $XY$ case with $\alpha=-1/2$ we like to note more
explicitly the qualitative result
\be\label{Imdiverge2}
\Im\chi(\omega\pm i\epsilon)=\cases{
\pm\frac{{\rm sign}(\omega)}
{\sqrt{\Lambda^2-\omega^2}},& for $|\omega|<\Lambda$,\cr
0,& for $|\omega|>\Lambda$.\cr}
\ee
with Kramers-Kronig transform
\be\label{Rediverge2}
\Re\chi(\omega)=\cases{
 \frac 2\pi\frac{{\rm arsinh}\sqrt{\left(\frac\Lambda\omega\right)^2-1}}
{\sqrt{\Lambda^2-\omega^2}},& for $|\omega|<\Lambda$,\cr
-\frac 2\pi\frac{{\rm arcsin}\sqrt{\frac\Lambda\omega}}
{\sqrt{\omega^2-\Lambda^2}},& for $|\omega|>\Lambda$.\cr}
\ee

For overcritical frustration\cite{CPK+95} $J_2>J_c=0.2411$ the
Heisenberg chain exhibits a gapped spectrum with a lower bound
$\Omega_g$. Considering square root divergences at the lower and upper
edge of the spectrum  
\be\label{overim}
\Im\chi(\omega\pm i\epsilon)=\cases{
\pm\frac{{\rm sign}(\omega)}
{\sqrt{(\omega^2-\Omega_g^2)(\Lambda^2-\omega^2)}},&  \hspace*{-3ex} for 
$\Omega_g<|\omega|<\Lambda$,\cr
0,& \hspace*{-3ex} else.\cr}
\ee
we obtain the Kramers-Kronig transform 
\be\label{overre}
\Re\chi(\omega)=\cases{
0,& for $\Omega_g<|\omega|<\Lambda$,\cr
\pm\frac{{\rm sign}(\omega)}
{\sqrt{(\omega^2-\Omega_g^2)(\omega^2-\Lambda^2)}},& else.\cr}
\ee
% In the simplified case of 
% \be
% \Im\chi(\omega\pm i\delta)=\cases{
% \pm\frac{{\rm sign}(\omega)}
% {\sqrt{(\omega-\Omega_g)(\Lambda-\omega)}},& for 
% $\Omega_g<|\omega|<\Lambda$,\cr
% 0,& else.\cr}
% \ee
% the Fourier transform can be given and reads
% \be\label{overeu}
% \chi(t)=2\pi\sin\left(\frac{\Lambda+\Omega_g}2t\right)
% J_0\left(\frac{\Lambda-\Omega_g}2t\right).
% \ee

The upper band edge singularities and the resulting algebraic
real-time asymptotics exist only at sufficiently low temperatures. 
At intermediate temperatures the upper limit of the continuum yields 
an anti-symmetrized Lorentzian contribution.
\be\label{Imlorenz}
\Im\chi(q,\omega) \simeq L_-(\phi) - L_+(\phi)\,,
\ee
where 
\be
L_{\pm} = \frac{\Gamma \cos \phi - (\Lambda\pm\omega)|\sin\phi|}
               {\Gamma^2 + (\Lambda\pm\omega)^2}\,.
\ee
Limiting $0\le\phi\le\pi/2$ the real part is simply given by 
\be\label{Relorenz}
\Re\chi(q,\omega) \simeq L_-(\phi-\pi/2) + L_+(\phi-\pi/2)
\ee
and the Fourier transform reads
\be\label{Eulorenz}
\chi (q,\omega) \simeq {\rm e}^{-\Gamma t}\ 
            \sin(\Lambda t + \phi)\,.
\ee
This temperature range will be referred to as "diffusive regime".

%%%%%%%%%%%%%%%%%%%%%%%%%%%%%%%%%%%%%%%%%%%%%%%%%%%%%%%%%%%%%%%%%%%

\subsection{$XY$ model}\label{sectionXYmodel}

We demonstrate the overlap of the accurate short time scale results
from the exact diagonalization of finite systems and the asymptotic
behavior accessible by field theory for an exactly solvable case, the
$XY$ model, where $J_2=J_z=0$. The spin operators in this model can be
transformed to non-interacting, spinless fermions via a Jordan-Wigner
transformation.\cite{Frad91} The structure factor Eq.\
(\ref{structure}) can be given for $L\to\infty$ in closed
form.\cite{FLS97} The imaginary part of the susceptibility at $q=\pi$
is 
\begin{equation}\label{XYLindom}
{\rm Im}\,\chi_{\rbx{XY}}(\pi,\omega) =
                \tanh(\beta\omega/4)\
                \left(4 - \omega^2\right)^{-0.5}
\end{equation}   
This is the field theoretical result Eq.\ (\ref{ChiCFT}) with scaling
dimension  $x=1$ multiplied with the square root divergence at the
upper band edge. The limit of $T\to 0$ is given by Eqs.\
(\ref{Imdiverge2}) and (\ref{Rediverge2}).

\begin{figure}[bt]
   \epsfxsize=0.48\textwidth
   \centerline{\epsffile{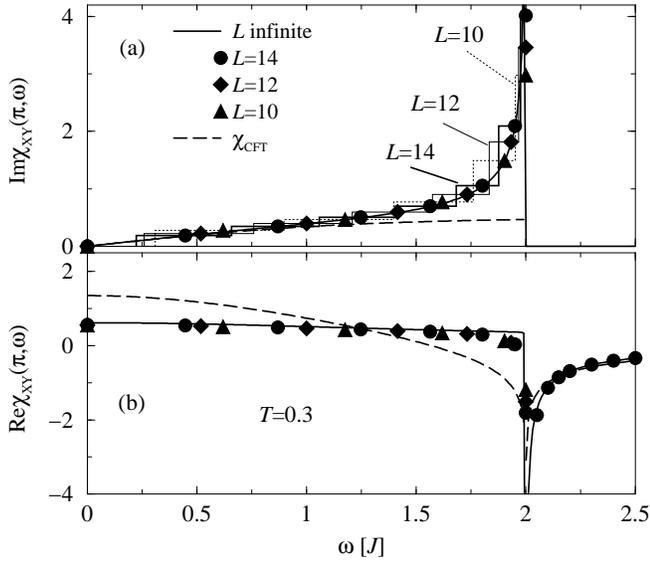}}
   \centerline{\parbox{\textwidth}{\caption{\label{FigzT3XY}
   \sl Susceptibility for the $XY$ model: (a) imaginary  part, (b) real
   part. The full line in (a) is the exact results from Eq.\
   (\protect\ref{XYLindom}), the symbols are obtained via Eq.\
   (\protect\ref{Imchidens}), and the step functions via Eq.\
   (\protect\ref{Imchibins}). The dashed line is the result from field
   theory with upper band edge cutoff but without divergence. The
   lines in (b) are the Kramers Kronig transforms of the imaginary
   part, and the symbols are obtained using Eq.\
   (\protect\ref{Rechiregular}) with $\Delta\omega=0$.}}} 
   \end{figure}

Fig.\ \ref{FigzT3XY}(a) shows the imaginary  part of
the susceptibility. The full line represents the exact results, the
symbols are obtained via the regularization Eq.\ (\ref{Imchidens}),
and the step functions are given by Eq.\ (\ref{Imchibins}). The dashed
line is the result from field theory with an UV-cutoff but without the
upper band edge divergence. Fig.\ \ref{FigzT3XY}(b) shows the real
part, the lines are the Kramers Kronig transforms of the imaginary
part, and the symbols are obtained using Eq.\ (\ref{Rechiregular})
with $\Delta\omega=0$.  

We conclude that the multiplicative approach of the low energy
description from field theory with the high energy behavior is
adequate. Also, the numerical approaches give a reasonable
approximation to the exact result. The values of the real part for
$\omega>\Lambda$ show only very little finite size effects. The
divergences of the real and the imaginary part at the upper band edge
show the correspondence predicted by Eqs.\ (\ref{Imdiverge2}) and
(\ref{Rediverge2}). 

The retarded, real-time correlation function can be determined
numerically in the thermodynamic limit.
\begin{equation}\label{XYLindeu}
\chi (q,t) = i \theta(t)
\lim_{L\to\infty}
\frac{1}{L}\sum_{k} \left(f_{k}-f_{k+q}\right)
         {\rm e}^{i(E_{k}-E_{{k}+{q}})t}
\end{equation}
The energy dispersion is given by $E_{k}=J\cos k$, $f_{k}$ are
Fermi distribution functions, and the sum covers the first Brillouin
zone. In general for $L\ge 10^4$ the result is independent of $L$ for
all practical purposes. 

   \begin{figure}[bt]
   \epsfxsize=0.48\textwidth
   \centerline{\epsffile{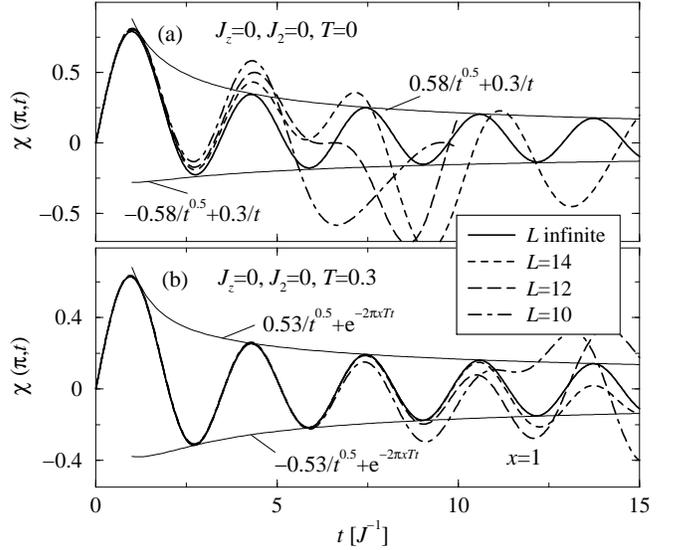}}
   \centerline{\parbox{\textwidth}{\caption{\label{FigtTXY}
   \sl Real-time spin-correlation function in the $XY$ case at $T=0$ (a)
   and $T=0.3$ (b). Thermodynamic limit results (full lines) are
   obtained via Eq.\ (\protect\ref{XYLindeu}). The different broken
   lines show the deviation of results for finite systems Eq.\
   (\ref{Realtimefinite}). $L=14$ yields a good representation of the
   correlation function up to $t \approx 2/J$ at $T=0$ and up to $t
   \approx 8/J$ at $T=0.3$. The thin lines show the asymptotic
   behavior as predicted in Sec.\ \protect\ref{sectionCFT}.}}}  
   \end{figure}

In Fig.\ \ref{FigtTXY} we show the retarded spin-correlation
function for finite systems compared with the result in the
thermodynamic limit ($L\to\infty$, full lines) at $T=0$ (a) and at
$T=0.3$ (b). The different broken lines show the deviation
of results for finite systems Eq.\ (\ref{Realtimefinite}). $L=14$
yields a good representation of the correlation function up to $t
\approx 2/J$ at $T=0$ and up to $t \approx 8/J$ at $T=0.3$. 

The thin solid lines in Fig.\ \ref{FigtTXY} show the asymptotic
$t^{-0.5}$ behavior from the upper band edge divergence. The
contribution from the low frequencies yields an additive term $\sim
t^{-1}$ at $T=0$ and $\sim {\rm e}^{-x2\pi Tt}$ with $x=1$ at $T=0.3$
as predicted in Sec.\ \ref{sectionCFT}. The fits have been obtained
for $80<tJ<100$ to assure the asymptotic limit. The numerical data for
finite systems and for $T \approx 0.3$ yield a good representation of
the correlation function in the thermodynamic limit up to time scales
that are already dominated by the asymptotic, large time scale
behavior.

%%%%%%%%%%%%%%%%%%%%%%%%%%%%%%%%%%%%%%%%%%%%%%%%%%%%%%%%%%%%%%%%%%%%%%

\subsection{Technical outline of the approach}

Renormalization group studies show that the $XY$ model is one point of
the line of critical fixed points towards which the interaction flows
in a bosonized representation of the Heisenberg model
($J_z=1$).\cite{Eme79,NO94} One thus expects qualitatively similar
results for the unfrustrated Heisenberg chain as in the $XY$
model. This should also hold for frustrated Heisenberg chains, at
least for under-critical $J_2\le J_c =
0.2411$.\cite{EN88,NO94,CPK+95}  

The discussion of the $XY$ model implies that the representation of the 
imaginary part of the correlation function is best achieved by
multiplying the upper band edge behavior to the field theoretical
expression.
\begin{equation}\label{CFTomfit}
{\rm Im}\,\chi(\pi,\omega) = 
      {\rm Im}[\chi_{\rbx{\rm CFT}}(\pi,\omega)]\,
                \frac{\left(\Lambda^2 - \omega^2\right)^{\alpha}}
                     {2\ \Lambda^{2\alpha}}\, \theta(\Lambda-|\omega|)
\end{equation}   
The real part of the susceptibility is given by the numerical Kramers
Kronig transform (KKT) of Eq.\ (\ref{CFTomfit}). The real-time
representation is obtained by the Fourier transform (FT) of
$\chi(\pi,\omega)$. We attempt this approach for frustrated
Heisenberg chains, where the exact form of the correlation function is
not known.

The field theoretical expression in Eq.\ (\ref{CFTomfit}) depends on
the parameter of the scaling dimension $x$ and a global prefactor
$v^{1-2x}\,\chi_{\rbx{0}}$. The latter is determined by requesting the
sum rule of the first moment of the susceptibility to be correctly
reproduced. The sum rule   
\begin{equation}\label{sumrule}
I_1(q,T) = \frac{1}{\pi}\int_{-\infty}^\infty d\omega\ \omega\ 
                          {\rm Im}\,\chi(q,\omega) 
\end{equation}   
can be extracted very accurately from the finite size data as
discussed in Sec.\ \ref{sectionSumrule}. We are then left with the
parameter vector ${\bf p}(T)=[x, \Lambda, \alpha]$ which we find to be
temperature dependent. The scaling variable $x$ determines the low
frequency behavior of the imaginary and the real part of the
susceptibility (Eq.\ (\ref{ChiTzero})) as well as the decay in real
time space as given by Eq.\ (\ref{timelimitCFT}). The upper  
continuum edge $\Lambda$ positions the cusp or divergence of the real 
part (Eq.\ (\ref{KKcutoff})) as well as the oscillatory behavior of
$\chi (\pi,t)$ as a function of time (Eq.\
(\ref{FTcutoff})). Finally the exponent $\alpha$ describes the shape
of the real part cusp or divergence (Eq.\ (\ref{KKcutoff})) and the
decay of the oscillations in time (Eq.\ (\ref{FTcutoff})).

%%%%%%%%%%%%%%%%%%%%%%%%%%%%%%%%%%%%%%%%%%%%%%%%%%%%%%%%%%%%%%%%%%%%
%%%%%%%%%%%%%%%%%%%%%%%%%%%%%%%%%%%%%%%%%%%%%%%%%%%%%%%%%%%%%%%%%%%%

\section{Sum rules and prefactors}\label{sectionSumrule}

Since the imaginary and real representation of the spin-correlation
function are Kramers Kronig related and as a consequence of the
bounded excitation spectrum\cite{boundspecquote} it is straight
forward to find that the sum rule Eq.\ (\ref{sumrule}) is given by  
\be\label{calcsumrule}
I_1(q,T) = -\lim\nolimits_{\omega\to\infty}\ \omega^2\ \Re\chi(q,\omega).
\ee
For $\lim_{T\to 0}I_1(q,T)=4K_1(q)$ the structure factor sum rule
discussed in Ref.\ \onlinecite{KMB+97} is reproduced. For arbitrary
frustration $J_2$ the structure factor sum rules are connected to the
ground state energy $E_G$ of the system via $K_1(\pi)/2 + K_1(0.5\pi) =
2E_G/3$. We recall that for $J_2=0$ and $J_2=0.5$ one has
$K_1(q)=2(1-\cos q)E_G/3$. For $J_2=0.5$ the average $\langle S^z_l
S^z_{l+2} \rangle_{T=0}=0$ vanishes.\cite{MG69} 

\begin{figure}[bt]
   \epsfxsize=0.48\textwidth
   \centerline{\epsffile{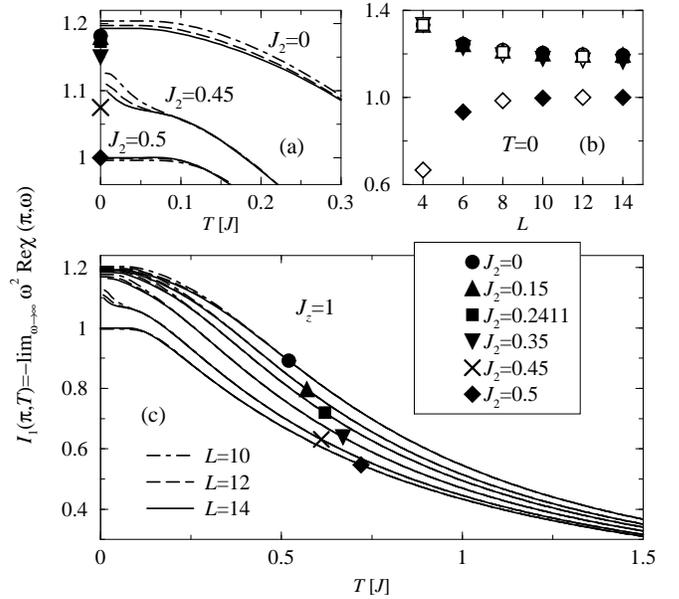}}
   \centerline{\parbox{\textwidth}{\caption{\label{ChilowT}
   \sl First moment of the spin susceptibility as extracted from the
   asymptotics of the real part in finite systems. (a) low temperature
   finite size effects and thermodynamic limit values as extracted
   from the finite size scaling represented in (b). (c) Temperature
   dependence.
   }}}
   \end{figure}

The crucial point is that $\Re\chi(q,\omega\to\infty)$ depends only
weakly on the system size $L$.\cite{Wern01a} Figure \ref{ChilowT}(c)
shows $I_1(q,T)$ as a function of temperature for different system
sizes and frustration parameters. In Fig.\ \ref{ChilowT}(a) it
becomes obvious that the result for $J_2=0$ and $L=14$ is for all
practical purposes in the thermodynamic limit for $T>0.3$. Analyzing
Eq.\ (\ref{realspaceCFT}) one finds the correlation length to be
$\xi=v/(2 \pi x T)$. For $x\sim 0.5$ and $v= 0.5
\pi(1-1.12J_2)$ we find that the finite size effects are of the order
of $10^{-3}$ when the correlation length becomes $\xi\sim L$.

Figure \ref{ChilowT}(b) shows the values of $I_1(\pi,0)$ as a
function of the system size. We determine the thermodynamic limit with
the algebraic scaling function $I_1(\pi,0,L)=I_1(\pi,0,\infty)+A_0\,
L^{-\eta}$. Systems with 
$L\, {\rm mod} \, 4=0$ (open symbols in Fig.\
\ref{ChilowT}(b)) 
in general converge differently than systems with $L\,
{\rm mod}\, 4=2$ (full symbols). The two cases yield two 
values the difference of which serves as an error estimate. For
$J_2=0$ we find $I_1(\pi,0,\infty)=1.1821(7)$ which is very close to the
exact value of $I_1(\pi,0)=1.1817258\ldots$. For $J_2=0.15$ and
$J_2=0.2411$ we find $I_1(\pi,0)=1.185(10)$ and $I_1(\pi,0)=1.173(2)$,
respectively. A value of $I_1(\pi,0)=1.1521(5)$ has been found for
$J_2=0.35$. The result for $J_2=0.5$ with $I_1(\pi,0)=1.000(3)$ is
extremely close to the exact value of 1. In Fig.\ \ref{ChilowT}(a)
the peculiar finite size effects for $J_2=0.45$ become
apparent. Obviously they result from the gap value\cite{CPK+95} of 
$\Omega_g\approx 0.12$ being just in the temperature range where the
finite size effects appear. The determination of
$I_1(\pi,0)=1.075(10)$ is thus less accurate. The results are represented
by the symbols in Fig.\ \ref{ChilowT}(a).

In Fig.\ \ref{ChilowT}(c) the temperature dependence of $I_1(\pi,T)$
is shown for different values of the frustration parameter. 
The general asymptotic behavior of $\lim_{T\to\infty} \chi(\pi,\omega)
\sim T^{-1}$ becomes apparent from the discussion in section
\ref{sectionHitemp}. For the first moment we find $\lim_{T\to\infty}
I_1(\pi,T) = 0.5/T$. This is reminiscent of the structure factor
sum rule  
\be\label{ChiThiom}
-\lim_{T\to\infty \atop \omega\to\infty}T\omega^2\Re\chi(\pi,\omega)=
\lim_{T\to\infty} \int\limits_{-\infty}^\infty d\omega'\, 
                                \omega'^2\, S(\pi,\omega')=0.5
\ee
and is generic for all values of $J_2$ and the $XY$ model. Note that
$\lim_{T\to\infty}S(q,\omega) = \lim_{T\to\infty}S(q,-\omega)$. 

\begin{figure}[bt]
   \epsfxsize=0.48\textwidth
   \centerline{\epsffile{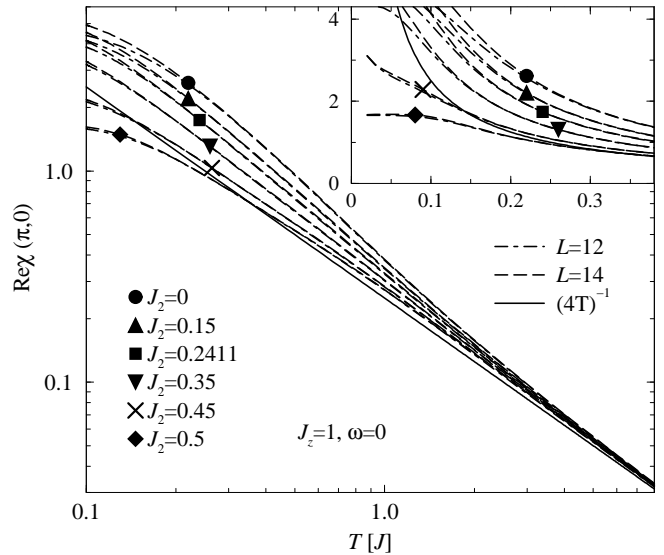}}
   \centerline{\parbox{\textwidth}{\caption{\label{Chinull}
   \sl Log-Log plot of $\Re\chi(\pi,\omega=0)$ from Eq.\
   (\protect\ref{Rechispectral}) as a function of temperature. The
   full line is the universal large $T$ asymptotic result
   $\sim0.25/T$. The inset shows the finite size effects at low
   temperatures. $\lim_{T\to 0}\Re\chi(\pi,\omega=0)$ diverges for
   $J_2\le 0.2411$ and saturates for $J_2 > 0.2411$.
   }}}
   \end{figure}

The values of $\Re\chi(\pi,\omega=0)$ show little finite size effects
even at rather low temperatures. Figure \ref{Chinull} shows a
Log-Log plot from Eq.\ (\ref{Rechispectral}) for different values of
the frustration $J_2$ and chain lengths as a function of temperature
(broken lines). The full line shows the asymptotic behavior 
\be\label{ChiTloom}
-\lim_{T\to\infty}T\,\Re\chi(\pi,0)=
\lim_{T\to\infty} \int_{-\infty}^\infty d\omega'\ S(\pi,\omega')=0.25
\ee
which reproduces a structure factor sum rule and is identical for all
values of the frustration and the $XY$ model. The field-theoretical
prediction Eq.\ (\ref{ChiCFT}) for the prefactor $T^{2x-2}$ with
constant scaling variable $x$ is clearly inappropriate for the
temperature range shown. Our analysis in Sec.\ \ref{sectionHeisen}
shows that the deviation results from an explicit temperature
dependence of the $x(T)$ as well as from the temperature dependence of
the singularity at the upper band edge, i.e., $\Lambda(T)$ and
$\alpha(T)$.

The inset of  Fig.\ \ref{Chinull} shows the finite size effects at low
temperatures. $\lim_{T\to 0}\Re\chi(\pi,\omega=0)$ diverges for
$J_2\le 0.2411$ and saturates for $J_2 > 0.2411$ which is reminiscent
of the presence of a gap.\cite{CPK+95}

\begin{figure}[bt]
   \epsfxsize=0.48\textwidth
   \centerline{\epsffile{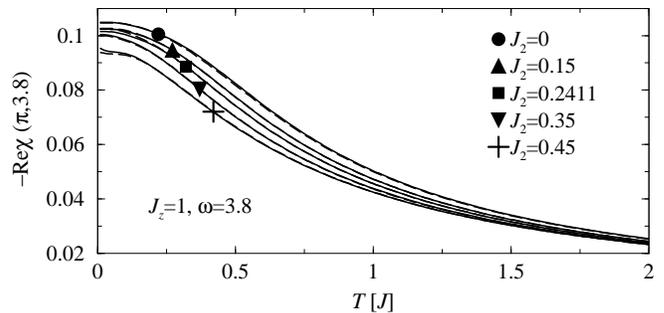}}
   \centerline{\parbox{\textwidth}{\caption{\label{Chi38}
   \sl Temperature dependence of $\Re\chi(\pi,3.8)$ from Eq.\
   (\protect\ref{Rechispectral}). The finite size effects are
   $\le0.1$\% and hardly visible on this scale (full lines $L=14$,
   broken lines $L=12$).
   }}}
   \end{figure}

For completeness we show in Fig.\ \ref{Chi38} the temperature
dependence of $\Re\chi(\pi,3.8)$ from Eq.\ (\ref{Rechispectral}) for 
different values of $J_2$. The finite size effects are 
$\le0.1$\% and hardly visible on this scale (full lines $L=14$, broken
lines $L=12$). We do not show plots for $J_2=0.5$ since the presence of
bound states makes the result unreliable, c.f.\ section
\ref{sectionOver}. 

%%%%%%%%%%%%%%%%%%%%%%%%%%%%%%%%%%%%%%%%%%%%%%%%%%%%%%%%%%%%%%%%%%%%
%%%%%%%%%%%%%%%%%%%%%%%%%%%%%%%%%%%%%%%%%%%%%%%%%%%%%%%%%%%%%%%%%%%%

\section{Frustrated Heisenberg chains}\label{sectionHeisen} 

We now turn to the determination of line shapes of the
spin-correlation function in frustrated Heisenberg chains making use
of the precise results obtained above.

\subsection{Critical frustration}\label{sectionCritical} 

We first discuss the values of $J_z=1$ and $J_2=J_c$ at the quantum
critical point making the field-theoretical results eligible for
comparison. In frequency space at $T=0$ for a 14 site chain there are
four spectral lines at frequencies $\tw_j \in {\cal W}^{(14)}_{.2411}
= \{0.264, 1.309, 2.112, 2.437\}$ which, by analogy to the dimer-dimer
correlation functions,\cite{Wern01a} may be identified as the triplet
excitations out of the ground state.\cite{degenquote} It is thus
reasonable to suppose them to form a well defined continuum in the
thermodynamic limit and thus Eqs.\ (\ref{Imchidens}) and
(\ref{Rechiregular}) can be applied with $\Delta\omega=0$.

\begin{figure}[bt]
   \epsfxsize=0.48\textwidth
   \centerline{\epsffile{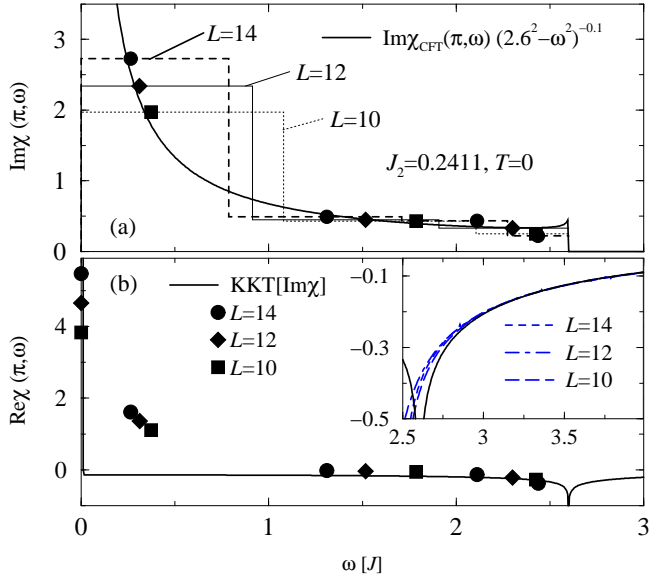}}
   \centerline{\parbox{\textwidth}{\caption{\label{FigzT0a241}
   \sl Imaginary part (a) and real part (b) of the spin-correlation
   function in the frustrated Heisenberg chain at $T=0$ 
   with $J_2=0.2411$. The imaginary part for finite systems is binned
   (Eq.\ (\protect\ref{Imchibins})), each bin holds one spectral line,
   symbols are from Eq.\ (\protect\ref{Imchidens}). The symbols for
   the real part are obtained by using Eq.\
   (\protect\ref{Rechiregular}). The full lines are the 
   theoretical result from Eq.\ (\protect\ref{CFTomfit}) and its
   KKT. Inset: enlargement of cutoff region with finite size results
   from  Eq.\ (\protect\ref{Rechispectral}).
   }}}
   \end{figure}

\begin{figure}[bt]
   \epsfxsize=0.48\textwidth
   \centerline{\epsffile{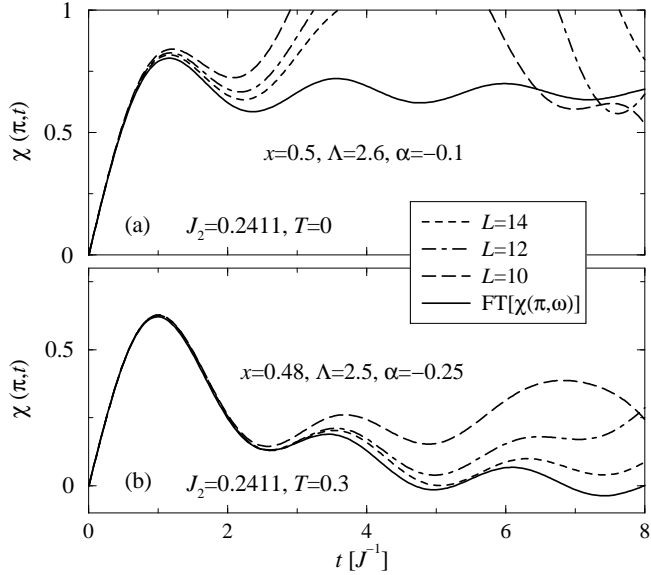}}
   \centerline{\parbox{\textwidth}{\caption{\label{FigtTa241}
   \sl Real-time spin-correlation function for $J_2=0.2411$ at
   (a) $T=0$ and (b) $T=0.3$. Broken lines are finite size data from
   Eq.\ (\protect\ref{Realtimefinite}), full lines are FT of Eq.\
   (\protect\ref{CFTomfit}). 
   }}}
   \end{figure}

The imaginary part of the spin-correlation function is shown in
Fig.\ \ref{FigzT0a241}(a). The bins are obtained using Eq.\
(\ref{Imchibins}) and the symbols are given using Eq.\
(\ref{Imchidens}). The symbols in Fig.\ \ref{FigzT0a241}(b) show the
real part as given by Eq.\ (\ref{Rechiregular}). For $\omega=0$ finite
size effects are  significant since one expects from field theory and
by analogy to the $XY$ model ${\rm
Re}\,\chi^{(T=0)}_{\rbx{}}(\pi,\omega=0) \to \infty$. For $\omega > 3$
the numerical results are essentially in the thermodynamic
limit as can be seen from the different broken lines in
the inset of Fig.\ \ref{FigzT0a241}(b). The real-time
representation of the spin-correlation function at $T=0$ is given in
Fig.\ \ref{FigtTa241}(a). As in the $XY$ case the system with $L=14$
yields a useful representation of the correlation function up to $t
\approx 2/J$.

The fit with the theoretical predictions from Eq.\ (\ref{CFTomfit}),
its KKT, and FT are given by the full lines in Figs.\ \ref{FigzT0a241}
(a), \ref{FigzT0a241}(b), and \ref{FigtTa241}(a) using the parameter
set ${\bf p}_{0.2411}(0)=[0.50(1),2.6(1),-0.10(7)]$. The cutoff
$2.5<\Lambda$ is bound by the highest spectral lines which
must lie in the continuum. Previous results\cite{YS97} suggest
$\Lambda$ to decrease monotonously with temperature 
and for reasons of consistency
$\Lambda<2.7$. The three parameters are then determined by matching
the first maximum as well as the slope for $1<tJ<2$ in the real
time representation and the value of the real part for $\omega\sim 3.8$
(c.f.~Fig.~(\ref{Chi38})). The finite size effects require to allow for
rather large error margins.

The result of $x$ is consistent with the prediction from field
theory. The value of $\alpha\neq 0$ suggests a more complicated upper
continuum edge than a simple ultraviolet cutoff. We emphasize that
the overall prefactor of the fit function is fixed by the sum rule Eq.\
(\ref{sumrule}) and that values for $\Lambda$ and $\alpha$ have been
obtained without using the not so well defined binned data of the
imaginary part.

The plot of the real-time representation of the spin-correlation
function at $T=0.3$ in Fig.\ \ref{FigtTa241}(b) reveals the temperature
dependence of the parameter vector ${\bf p}_{0.2411}(0.3)=[0.48(1),
2.5(1), -0.25(5)]$. The result for $L=14$ yields a useful
representation of the correlation function up to $t \approx 4/J$. The
full line is the fit from the FT of Eq.\ (\ref{CFTomfit}). The
exponential fall off predicted in Sec.\  \ref{sectionCFT} is confirmed
and renders the value of the scaling dimension. The oscillations are
much less damped than would be obtained with an upper band edge
exponent of $\alpha=0$ thus yielding the negative value of
$\alpha=-0.25$. The cutoff $\Lambda$ is given via the period of the
oscillations.  

   \begin{figure}[bt]
   \epsfxsize=0.48\textwidth
   \centerline{\epsffile{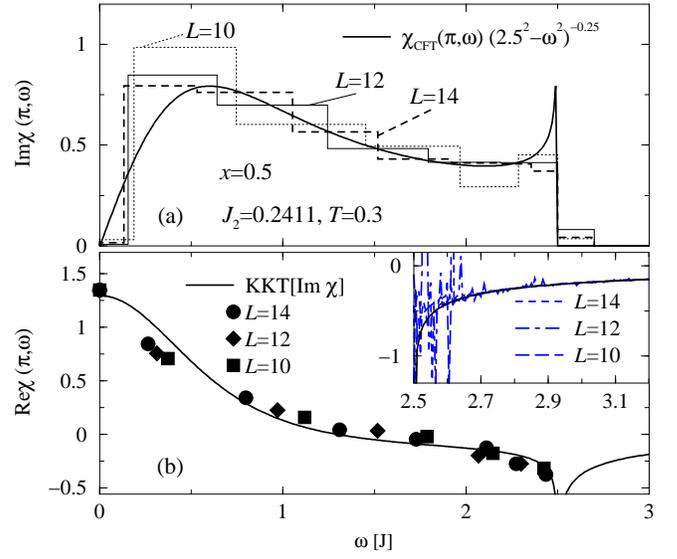}}
   \centerline{\parbox{\textwidth}{\caption{\label{FigzT3a241}
   \sl Imaginary part (a) and real part (b) of the spin-correlation
   function in the frustrated Heisenberg chain at $T=0.3$ 
   with $J_2=0.2411$. The imaginary part for finite systems is binned
   (Eq.\ (\protect\ref{Imchibins})), the symbols for the real part are
   obtained using Eq.\ (\protect\ref{Rechiregular}). The full lines
   are the theoretical result from Eq.\ (\protect\ref{CFTomfit}) and its
   KKT. Inset: enlargement of cutoff region with finite size results
   from  Eq.\ (\protect\ref{Rechispectral}).   
   }}}
   \end{figure}

Fig.\ \ref{FigzT3a241}(a) and \ref{FigzT3a241}(b) show the imaginary
and real part of 
the frequency representation of the spin-correlation function at
$T=0.3$, respectively. The binned data for the imaginary part via Eq.\
(\ref{Imchibins}) and the symbols for the real part via Eq.\
(\ref{Rechiregular}) are obtained for the sets of dominant spectral
lines, in the case of $L=14$ they are given by the frequencies $\tw_j
\in \tilde{\cal W}^{(14)}_{.2411} = \{0.311, 0.971, 1.517, 2.068,
2.301\}$. The condition of a well defined continuum with respect to
Bethe ansatz quantum numbers is violated and thus Eq.\
(\ref{Imchidens}) cannot be applied any more.  For the real part the
data are regularized with $\Delta\omega=0.1$, which is determined to
give reliable results analogously to the dimer-dimer correlation
functions.\cite{Wern01a} An exception is made at $\omega=0$, where no
regularization is applied ($\Delta\omega=0$).

The good correspondence of the field-theoretical fits from Eq.\
(\ref{CFTomfit}) and its KKT (solid lines) in Figs.\ \ref{FigzT3a241}
(a) and \ref{FigzT3a241}(b) proves the reliability of the parameters
extracted from the real-time representation. Especially the good
agreement of 
the values of $\Re\chi(\pi,0)$ and of $\Re\chi(\pi,\omega>2.7)$
(inset Fig.\ \ref{FigzT3a241}(b)) are non-trivial consistency
checks. We expect a thermal smearing out of the small divergence at 
the upper band edge of which the shape is not known and which we did
not account for (solid line in Fig.\ \ref{FigzT3a241}(a)). This might
have a small influence on the parameters extracted and thus we
adapted rather conservative error bars. The finite size data in both
the real and imaginary part suggest a steeper slope in the low
frequency dependence of the fitted curves. Together with the
temperature dependence of the scaling dimension $x$ this indicates the
breakdown of the scale invariance predicted by field theory at finite
temperatures.\cite{lowomegaquote}

%%%%%%%%%%%%%%%%%%%%%%%%%%%%%%%%%%%%%%%%%%%%%%%%%%%%%%%%%%%%%%%%%%%%%%%

\subsection{Unfrustrated Heisenberg chain}

Heisenberg chains without frustration are relevant for most of the
magnetically quasi one-dimensional systems studied
experimentally. Since the system is integrable the numerical data can
be compared to results from Bethe ansatz.

At $T=0$ the four spectral lines of the triplet excitations out of the
ground state for a 14 site chain are at frequencies $\tw_j \in {\cal
W}^{(14)}_{0} = \{0.307, 1.57, 2.56, 3.10\}$.\cite{KM97,KHM98,KM00} It
is thus reasonable to suppose them to form a well defined continuum in
the thermodynamic limit and thus Eqs.\ (\ref{Imchidens}) and
(\ref{Rechiregular}) can be applied with $\Delta\omega=0$. Binned data
for the imaginary part are obtained via Eq.\ (\ref{Imchibins}).

\begin{figure}[bt]
   \epsfxsize=0.48\textwidth
   \centerline{\epsffile{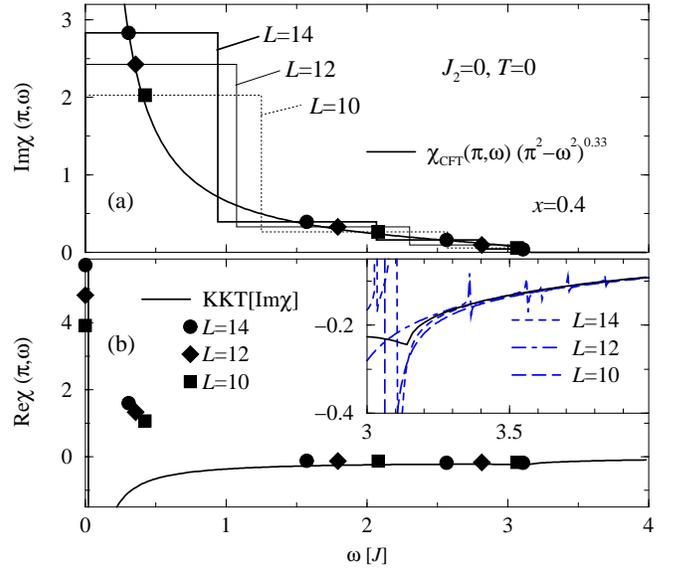}}
   \centerline{\parbox{\textwidth}{\caption{\label{FigzT0a000}
   \sl Imaginary part (a) and real part (b) of the spin-correlation
   function in the unfrustrated Heisenberg chain at $T=0$. The binned
   curves in (a) are from (Eq.\ (\protect\ref{Imchibins})), the
   symbols are from Eq.\ (\protect\ref{Imchidens}). The symbols for 
   the real part (b) are obtained using Eq.\
   (\protect\ref{Rechiregular}). The full lines are the fits from
   Eq.\ (\protect\ref{CFTomfit}) and its KKT. Inset: enlargement of
   cutoff region with finite size results from  Eq.\
   (\protect\ref{Rechispectral}).}}} 
   \end{figure}

\begin{figure}[bt]
   \epsfxsize=0.48\textwidth
   \centerline{\epsffile{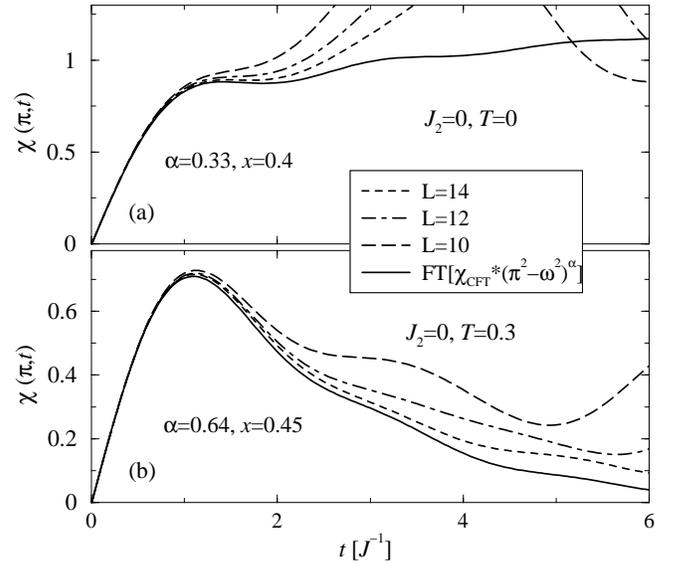}}
   \centerline{\parbox{\textwidth}{\caption{\label{FigtTa000}
   \sl Real-time spin-correlation function for $J_2=0$ at
   (a) $T=0$ and (b) $T=0.3$. Broken lines are finite size data from
   Eq.\ (\protect\ref{Realtimefinite}), full lines are FT of Eq.\
   (\protect\ref{CFTomfit}).}}}
   \end{figure}

The imaginary part of the spin-correlation function is shown in
Fig.\ \ref{FigzT0a000}(a). The real part in Fig.\ \ref{FigzT0a000}
(b) shows for $\omega=0$ significant finite size effects since ${\rm
Re}\,\chi^{(T=0)}_{\rbx{}}(\pi,\omega=0) \to \infty$. For $\omega >
3.5$ the numerical results are essentially in the thermodynamic limit
(Inset of Fig.\ \ref{FigzT0a000}(b)). The real-time representation of
the spin-correlation function at $T=0$ is given in Figure
\ref{FigtTa000}(a).

The upper edge of the two-spinon continuum is know exactly to be
$\lambda=\pi$.\cite{dCP62} Bethe ansatz results suggest that the infra
red divergence of the two-spinon contribution $\chi^{(2)}$ to the
imaginary part of the spin-correlation function has a logarithmic
correction  
\be
\lim\nolimits_{|\omega| \to 0}\Im\chi^{(2)}(\pi,\omega) \propto
   \omega^{-1}\sqrt{\ln(\omega^{-1})}
\ee 
while at the upper continuum edge it vanishes square root
like.\cite{KMB+97} The two-spinon contribution has been found to
contribute 72.89\% to the total spectral weight.\cite{KMB+97} Our and
previous numerical studies show that the spectral weight of the total
correlation function for above the two-spinon continuum ($\omega>\pi$)
at $q=\pi$ is less than 0.1\%.\cite{FLS97,YS97} Thus the total
spin-correlation function includes also about $27$\% higher order
contributions and we have 
$\Im\chi(\pi,\omega)>\Im\chi^{(2)}(\pi,\omega)$. Consequently we must
require $x\le 0.5$, $\alpha\le 0.5$, and $\Lambda=\pi$.  

From the amplitude in the real-time representation in Fig.\
\ref{FigtTa000} (a) we find that the parameters $x$ and $\alpha$ fall
on a line defined by $(0.38,0.25) < (x,\alpha) < (0.44,0.5)$. Taking
also the value of the real part at $\omega=4$ in the inset of  Fig.\
\ref{FigzT0a000}(b) into consideration we determine ${\bf
p}_0(0)=[0.40(3),\pi\pm0.01,0.33(5)]$. The error margins have been
chosen rather large because of the obvious finite size effects.
The resulting fits with the theoretical predictions from Eq.\
(\ref{CFTomfit}), its KKT, and FT are given by the full lines 
Figs.\ \ref{FigzT0a000}(a), \ref{FigzT0a000}(b), and \ref{FigtTa000}
(a) and show satisfactory agreement with the results from finite
systems.

   \begin{figure}[bt]
   \epsfxsize=0.48\textwidth
   \centerline{\epsffile{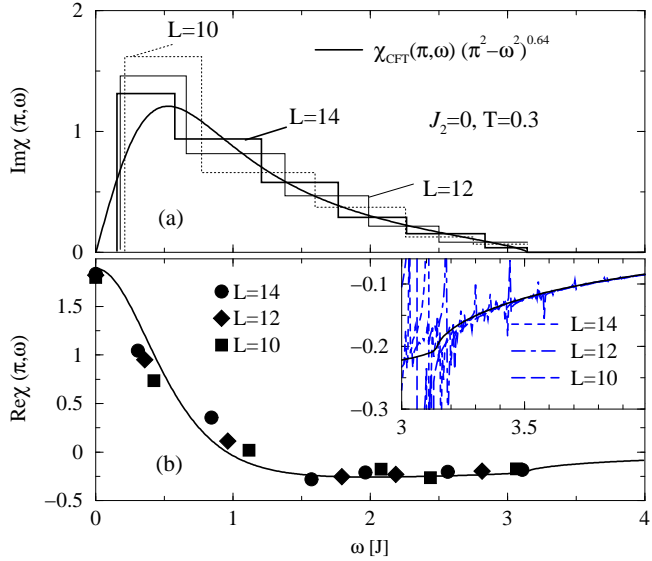}}
   \centerline{\parbox{\textwidth}{\caption{\label{FigzT3a000}
   \sl Imaginary part (a) and real part (b) of the spin-correlation
   function in the unfrustrated Heisenberg chain at $T=0.3$. The step
   function in (a) is from (Eq.\ (\protect\ref{Imchibins})), the
   symbols in (b) from Eq.\ (\protect\ref{Rechiregular}). The full lines
   are from Eq.\ (\protect\ref{CFTomfit}) and its KKT. Inset:
   enlargement of cutoff region with results from  Eq.\
   (\protect\ref{Rechispectral}).}}} 
   \end{figure}

The short-dashed line in Fig.\ \ref{FigtTa000}(b) shows the
correlation function for 14 sites in the real-time
representation. The exponential decay in time imply a scaling
dimension of $x>0.4$ and the strongly damped oscillations an exponent
$\alpha>0.5$. The real part at higher frequencies as shown in the
inset of  Figs.\ \ref{FigzT3a000}(b) requires the exponent to be
$\alpha<0.7$. Together with a matching value for
$\Re\chi(\pi,\omega=0)$ the parameter set ${\bf
p}_0(0.3)=[0.45(2), \pi\pm0.1, 0.64(10)]$ yields the best fits
from Eq.\ (\ref{CFTomfit})as shown by the full lines in Figs.\
\ref{FigzT3a000}(a), \ref{FigzT3a000}(b), and \ref{FigtTa000}
(b). The overall agreement is satisfactory, for small frequencies the
numerical data suggest a slightly altered functional dependence on the
frequency than is reproduced by the field-theoretical
fit.\cite{lowomegaquote}

%%%%%%%%%%%%%%%%%%%%%%%%%%%%%%%%%%%%%%%%%%%%%%%%%%%%%%%%%%%%%%%%%%%%%%%

\subsection{Overcritical frustration}\label{sectionOver}

In the case of overcritical frustration the spectrum of the spin
chains acquires a gap $\Omega_g$.\cite{MG69,CPK+95} We discuss here the
value of $J_2=0.5$ for better comparability with results from
literature.\cite{MG69,SS81,Uhri99} For a two particle (spinon)
continuum one expects a density of states that diverges square root
like both at the lower as well as at the upper edge. Previous
numerical\cite{YS97} and variational\cite{Uhri99} results suggest
sharp maxima in the density of states just above the lower edge
$\Omega_g$ and just below the upper edge $\Lambda$ of the continuum
accompanied by a square root like vanishing at both edges. This can be
understood in connection with bound states being present close to the
edge of the continuum.\cite{YMV96}

\begin{figure}[bt]
   \epsfxsize=0.48\textwidth
   \centerline{\epsffile{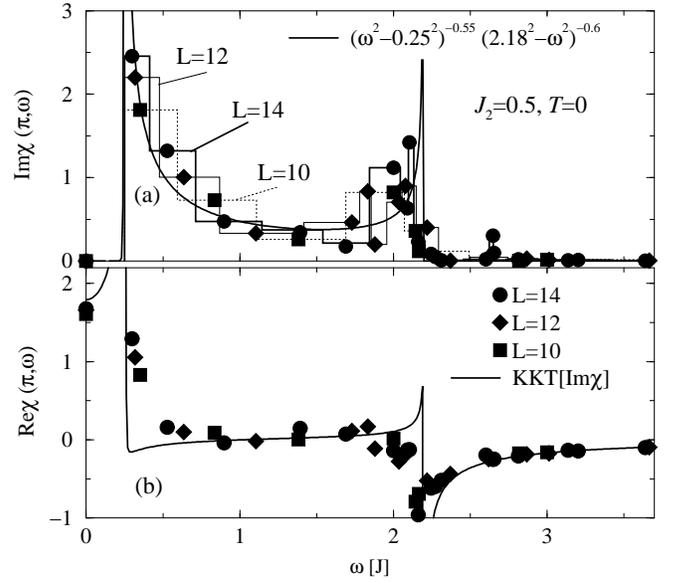}}
   \centerline{\parbox{\textwidth}{\caption{\label{FigzT0a500}
   \sl  Imaginary part (a) and real part (b) of the spin-correlation
   function for $J_2=0.5$ at $T=0$. The binned curves in (a) are from
   (Eq.\ (\protect\ref{Imchibins})), the symbols are from Eq.\
   (\protect\ref{Imchidens}). The symbols for the real part (b) are
   obtained using Eq.\ (\protect\ref{Rechiregular}). The full lines in
   are the fits from Eq.\ (\protect\ref{CFTomfit}) and its KKT.}}}
   \end{figure}

At $T=0$ there are more spectral lines present than in the case of
critical and undercritical frustration. Some of them are signatures of
the bound states present in the system.\cite{Uhri99} We still apply
Eqs.\ (\ref{Imchidens}) and (\ref{Imchibins}) to extract the imaginary
part of the spin-correlation function as shown in Fig.\
\ref{FigzT0a500}(a). The fluctuations of the results at the upper
band edge are reminiscent of the fact that the bound states do not form
a continuum in the thermodynamic limit. We do not attempt to refine
the plot by extracting the bound state contributions since results of
the real part and especially the real-time representation are
more reliable anyway. 

\begin{figure}[bt]
   \epsfxsize=0.48\textwidth
   \centerline{\epsffile{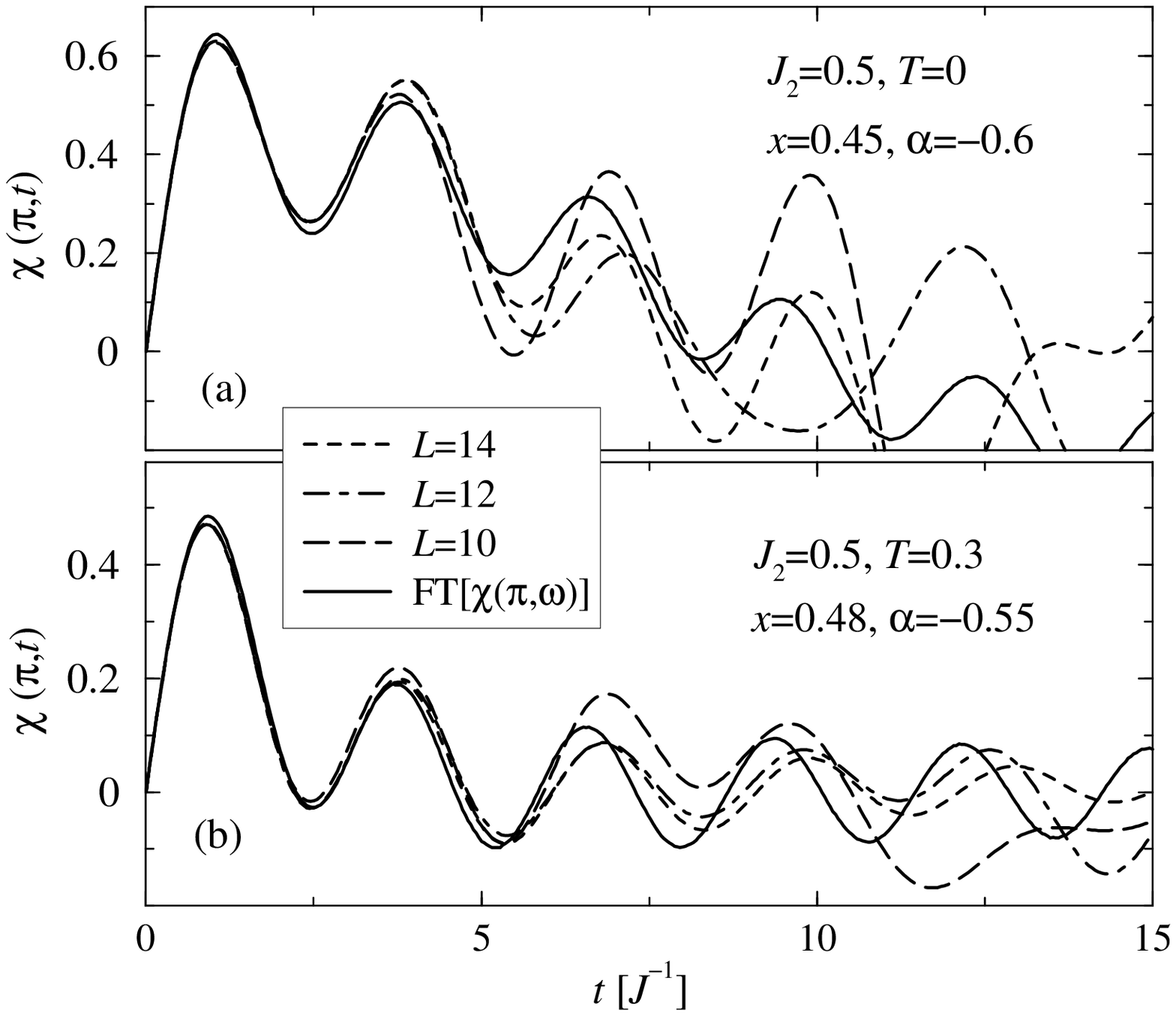}}
   \centerline{\parbox{\textwidth}{\caption{\label{FigtTa500}
   \sl Real-time spin-correlation function for $J_2=0.5$ at
   (a) $T=0$ and (b) $T=0.3$. Broken lines are finite size data from
   Eq.\ (\protect\ref{Realtimefinite}), full lines are FT of Eq.\
   (\protect\ref{CFTomfit}).}}}
   \end{figure}

Fig.\ \ref{FigzT0a500}(b) shows the real part from Eq.\
(\ref{Rechiregular}) with $\Delta\omega=0.1$ for $1.5<\omega<2.3$ and
$\Delta\omega=0.001$ else. The real-time representation of the
spin-correlation function at $T=0$ is given in Fig.\
\ref{FigtTa500}(a). 

In order to adapt Eq.\ (\ref{CFTomfit}) to the gapped spectrum we have
to include the value of the gap energy into the functions $F_x$ in
the field-theoretical expression Eq.\ (\ref{ChiCFT}) as
\be
F_x\left(\frac{\omega\pm v(q-\pi)}{2\pi T}\right)\ \to\
F_x\left(\frac{\omega\pm\sqrt{\Omega_g^2 + v^2(q-\pi)^2}}{2\pi
T}\right). 
\ee
At $T=0$ this yields the correct divergence at the lower band edge
$\Im\chi(\pi,\omega)\sim \left[\omega^2 - \Omega_g^2\right]^{x-1}$,
c.f.\ Eq.\ (\ref{ChiTzero}), while for finite temperatures
$\Im\chi(\pi,\omega)\sim \omega$. 

The full line in Fig.\ \ref{FigtTa500}(a) shows the fit from the FT 
of Eq.\ (\ref{CFTomfit}) with $x=0.45(3)$, $\Omega_g=0.25$,\cite{SS81}
$\Lambda=2.18(10)$, and $\alpha=-0.6(1)$. The slow damping of the
oscillations requires the large value of $|\alpha|$, there is a small
frequency modulation stemming from the lower boundary $\Omega_g$. The
agreement with the finite size data (broken lines) for the finite size
effect free time domain $tJ < 5$ is not nearly as good as for the
cases of lower frustration. We conclude that the simple functional 
form of Eq.\ (\ref{CFTomfit}) is insufficient. The square root
dependence reported in literature\cite{Uhri99} must be
included. A perturbative examination for small $J_z$ and $J_2$
shows\cite{YMV96} the strong interplay between the spinon continuum
and bound states. Since a theoretical expression for the correct
spectral weight distribution is not known we limit ourselves here to
the observation that the small damping of the oscillations in Fig.\
\ref{FigtTa500}(a) requires a rather sharp increase in
$\Im\chi(\pi,\omega\to\Lambda)$.  

The full lines in Fig.\ \ref{FigzT0a500}(a) and \ref{FigzT0a500}(b)
show the fit of Eq.\ (\ref{CFTomfit}) and its KKT to the
imaginary-part and real-part data, respectively. They are similar to
the forms given by Eqs.\ (\ref{overim}) and (\ref{overre}) where
$x=0.5$ and $\alpha=-0.5$. Here the parameters $x$ and $\alpha$ have
been adapted to roughly match the numerical values of $\Re\chi(\pi,0)$
and $\Re\chi(\pi,3.8)$. Especially the discrepancies of the fits at 
$\omega\sim\Lambda$ indicate that a more involved fit function is
necessary.  

\begin{figure}[bt]
   \epsfxsize=0.48\textwidth
   \centerline{\epsffile{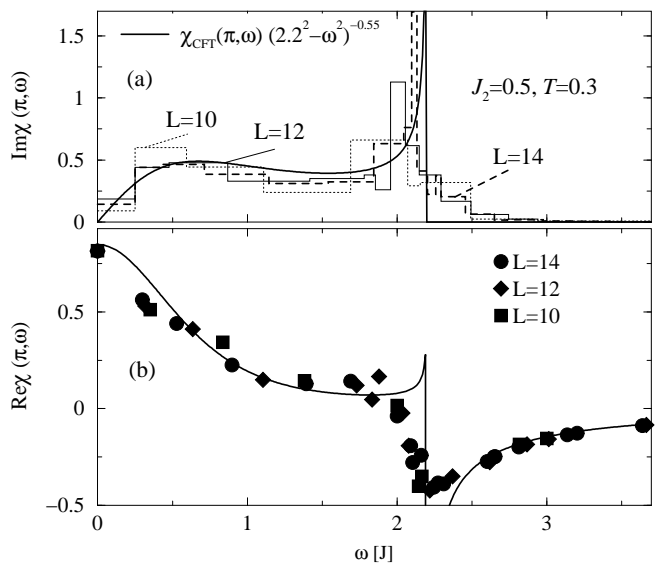}}
   \centerline{\parbox{\textwidth}{\caption{\label{FigzT3a500}
   \sl Imaginary part (a) and real part (b) of the spin-correlation
   function for $J_2=0.5$ at $T=0.3$. The step function in (a) is from
   (Eq.\ (\protect\ref{Imchibins})), the symbols in (b) from Eq.\
   (\protect\ref{Rechiregular}). The full lines are from Eq.\
   (\protect\ref{CFTomfit}) and its KKT.}}}
   \end{figure}

Fig.\ \ref{FigtTa500}(b) shows that also at $T=0.3$ an accurate fit
(full line) with the functional form of Eq.\ (\ref{CFTomfit}) to the
finite size data (broken lines) is not possible. The parameters for
the approximate fit are $x=0.48$, $\Omega_g=0.25$,\cite{SS81}
$\Lambda=2.2$, and $\alpha=-0.55$. Correspondingly, the fits to the
imaginary- and the real-part representations in Fig.\
\ref{FigzT3a500}(a) and \ref{FigzT3a500}(b) show inconsistencies with
the numerical data.

%%%%%%%%%%%%%%%%%%%%%%%%%%%%%%%%%%%%%%%%%%%%%%%%%%%%%%%%%%%%%%%%%%%%%%

\subsection{Intermediate temperatures}\label{sectionMitemp}

At intermediate temperatures the interaction in the system is expected
to broaden out all sharp features in the correlation functions. The
onset of this effect is already observed at $T=0.3$ as discussed in
the previous sections. At $T=0.7$ the exponent of the upper continuum
edge for $J_2=0$ is $\alpha=3.2(1)$ so that the singularity is
basically completely damped out. Also, the rather large effective
continuum edge $\Lambda=3.45(10)$ does not quite reproduce the correct
oscillatory behavior as a function of time. 

At about the same temperature the scaling dimension increases to $x\sim
1$. $T^*\approx 0.7$ thus marks the crossover temperature from
strongly interacting, conformally invariant to noninteracting fermion
and high energy diffusive behavior. This is consistent with
$\partial_T \Re\chi(\pi,\infty)$ and $\partial_T \Re\chi(\pi,3.8)$
being extremal at $T\approx T^*$ as seen in  Figs.\ \ref{ChilowT} and
\ref{Chi38}.

\begin{figure}[bt]
   \epsfxsize=0.48\textwidth
   \centerline{\epsffile{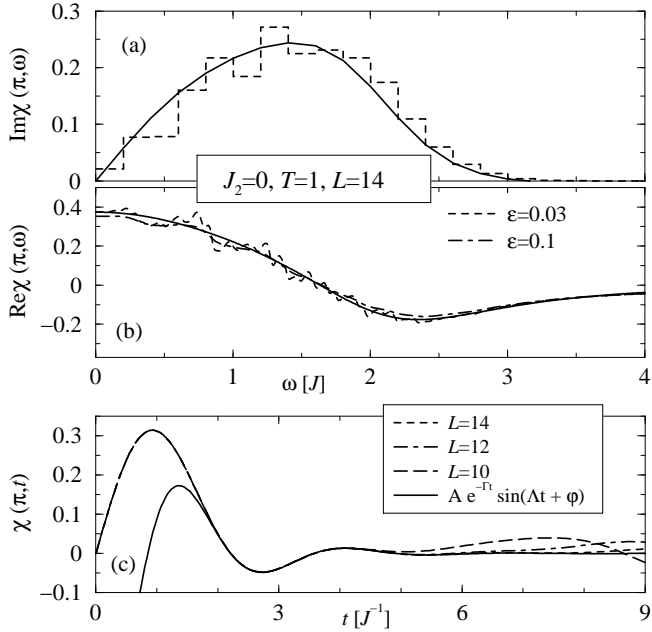}}
   \centerline{\parbox{\textwidth}{\caption{\label{FigT1a000}
   \sl Spin-correlation function of the unfrustrated Heisenberg chain
   at $T=1$. The broken line in (a) is the binned imaginary
   representation, the broken lines in (b) are the real part from Eq\
   (\ref{Rechispectral}) with two values of $\epsilon$ for
   regularization, all for $L=14$. Full lines in (a) and (b) are
   double Lorentzian fits. (c) shows the real-time representation
   from finite systems (broken lines) and the asymptotic fit (full
   line) with $\Lambda=2.31(1)$, $\Gamma=0.941(5)$, and $A=-0.67(5)$.}}} 
   \end{figure}

Figure \ref{FigT1a000}(c) shows the real-time representation of
the spin-correlation function of the unfrustrated Heisenberg chain at
$T=1$. The amplitude of the modulations between $tJ\approx 2$ and the
onset of finite size effect for the 14 site chain at $tJ\sim 7$ cannot
be fitted algebraically. The exponential fit from Eq.\
(\ref{Eulorenz}) shown by the full line in Fig. \ref{FigT1a000}(c)
matches excellently. The analogy to the $XY$ model suggests that the
long time asymptotics is captured in this fit. The extracted value for
$\Lambda=2.31(1)$ thus marks the effective, thermally smeared out
upper band edge. For the appropriate interpretation of the parameter
$\Gamma=0.941(5)$ refer to section \ref{sectionresults}.

The discrepancy of the real-time fit function (full line in
Fig. \ref{FigT1a000}(c)) and the correct line shape for small times
does not allow for a direct comparison of the results with its
Fourier transforms. The difference between the fit and the exact
result is roughly exponential. The fit with Eq.\ (\ref{Eulorenz})
implies that the real and imaginary part should contain contributions
from the continuum boundary Lorentzians Eqs.\ (\ref{Relorenz}) and
(\ref{Imlorenz}). Indeed, the double Lorentzian fits (full lines in
Fig. \ref{FigT1a000}(a) and \ref{FigT1a000}(b)) with an additive
Lorentzian contribution centered at $\omega=0$ compare well with the
binned data  for the imaginary part and the real part data from Eq.\
(\ref{Rechispectral}) with $\epsilon=0.03$ (broken lines). The fit
parameters even though similar are not such that the fits are
appropriately Kramers-Kronig and Fourier related. The fits must thus
be regarded as sophisticated guides to the eye. Similar results
are obtained for $J_2>0$. Similar line shapes are also found in
systems with large spins.\cite{Vill74}

%%%%%%%%%%%%%%%%%%%%%%%%%%%%%%%%%%%%%%%%%%%%%%%%%%%%%%%%%%%%%%%%%%%%%%

\subsection{High temperature limit}\label{sectionHitemp}

\begin{figure}[bt]
   \epsfxsize=0.48\textwidth
   \centerline{\epsffile{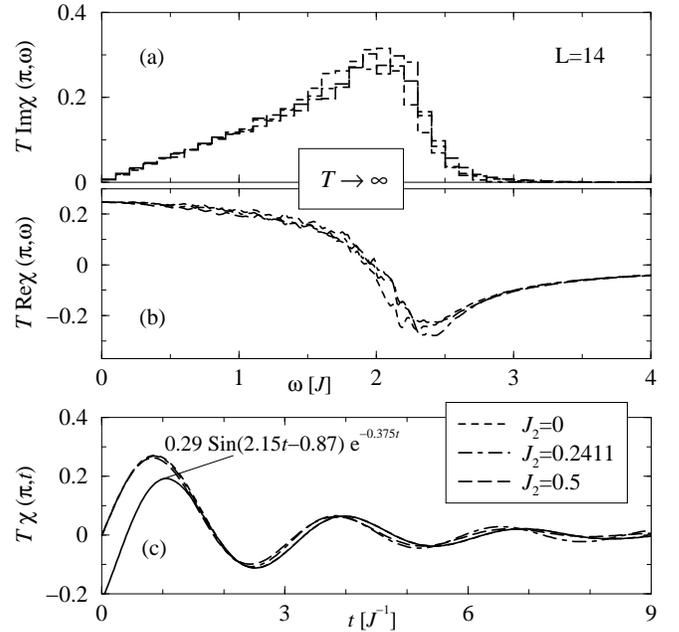}}
   \centerline{\parbox{\textwidth}{\caption{\label{FigtHighT}
   \sl Imaginary (a), real (b), and real-time (c) representation of
   the susceptibility for $T\to\infty$. Broken lines in (a) are binned,
   in Eq.\ (\protect\ref{Rechispectral}) with $\epsilon=0.02$ was
   applied, and in (c) Eq.\ (\protect\ref{Realtimefinite}). The solid
   line in (c) is the sinusoidal fit for the example of $J_2=0$.}}} 
   \end{figure}

In Fig.\ \ref{FigtHighT}(a), \ref{FigtHighT}(b), and
\ref{FigtHighT}(c) we show the respective imaginary, real, and
real-time representation of the susceptibility for different values of
the frustration in the limit of infinite temperatures. The data of the
imaginary part (a) are binned and the real part (b) is given by Eq.\
(\ref{Rechispectral}) with $\epsilon=0.02$ for $L=14$. The time
representations (c) from Eq.\ (\ref{Realtimefinite}) show finite size
effects for $t>6/J$.

For $\omega\to 0$ the slopes of the imaginary part are
similar to the exact result for the $XY$ model Eq.\
(\ref{XYLindom}). A small frustration dependence becomes obvious when
plotting the structure factor instead of the correlation
function.\cite{FLS97,FL98}

The oscillations in time shown in  Fig.\ \ref{FigtHighT}(c) can be
fitted very accurately for $1.5<tJ<6$ with an exponential decay via Eq.\
(\ref{Eulorenz}). The parameter sets $[\Lambda,\Gamma,\phi]$ are
obtained as $[2.15(1),0.375(2),-0.87(1)]$ for $J_2=0$,
$[2.31(1),0.323(2),-1.13(1)]$ for $J_2=0.2411$, and
$[2.21(1),0.345(2),-0.82(1)]$ for $J_2=0.5$. The full line shows the
resulting fit function for $J_2=0$.

In the classical limit, where $\langle S_z^2 \rangle_{T\to\infty} \to
\infty$, paramagnetic behavior is expected for $T\gg J$.
This leads to an expected functional dependence of the structure
factor of $\lim_{T\to\infty} S_{\rm class}(q,\omega)\sim
\lim_{\epsilon\to0}\, \epsilon/(\omega^2 + \epsilon^2)$. From Eq.\
(\ref{ChiTloom}) follows that $\lim_{T\to\infty} T\Re\chi(\pi,0) =
\langle S_z^2 \rangle_{T\to\infty}$ which is consistent with the
expected functional dependence in the classical limit. 
% A generalized form
% of Eq.\ (\ref{Imlorenz}). It is obtained by introducing general
% amplitudes $\sin\phi\to A_1$ and $\cos\phi\to A_2$. A fit to the
% umerical data  (broken lines) in Fig.\ \ref{FigtHighT}(c) is still
% not possible. 

The $XY$ model is one point of the line of critical fixed points
towards which the interaction flows in a bosonized representation of
the Heisenberg model.\cite{Eme79,NO94} The susceptibility of the $XY$
model shows a square root divergence at $\omega=2$ and
$\Im\chi_{\rbx{XY}}(\pi,\omega>2) \equiv 0$. The shape of the spectrum
at the upper band edge observed for Heisenberg chains is thus an
interaction effect.\cite{FLS97} The shape of
$\Im\chi(\pi,\omega\sim\Lambda)$ indeed resembles that of a Fermi
distribution of weakly interacting electrons. We thus interpret the
limit $T\to\infty$ as best described by weakly interacting spinless
fermions.

%%%%%%%%%%%%%%%%%%%%%%%%%%%%%%%%%%%%%%%%%%%%%%%%%%%%%%%%%%%%%%%%%%%%
%%%%%%%%%%%%%%%%%%%%%%%%%%%%%%%%%%%%%%%%%%%%%%%%%%%%%%%%%%%%%%%%%%%%

\section{Results}\label{sectionresults}

\begin{figure}[bt]
   \epsfxsize=0.48\textwidth
   \centerline{\epsffile{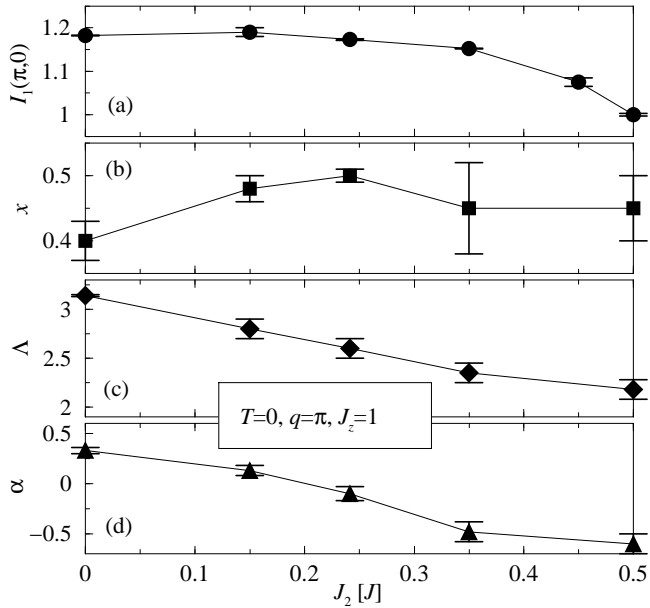}}
   \centerline{\parbox{\textwidth}{\caption{\label{ResultsJ2}
   \sl Extracted values for (a) the sum rule $I_1(\pi,0)$, 
   (b) the scaling dimension $x$, (c) the upper edge of the
   continuum, and (d) the exponent at the upper edge of the
   continuum as a function of $J_2$ at $T=0$.}}}  
   \end{figure}

Figure \ref{ResultsJ2} summarizes as a function of the frustration
$J_2$ at $T=0$ the extracted values for (a) the sum rule $I_1(\pi,0)$,
(b) the scaling dimension $x$, (c) the upper edge of the continuum
$\Lambda$, and (d) the exponent at the upper edge of the continuum
$\alpha$.

(a) The values of $I_1$ for $J_2\le 0.2411$ are within error bars almost
identical which underlines the common features of Heisenberg chains
with undercritical frustration.\cite{EN88,CPK+95,Eme79,NO94}

(b) The scaling variable $x$ shows a stronger infrared divergence for 
unfrustrated Heisenberg chains than for those with critical
frustration. The values for overcritical frustration have to be
regarded as effective ones as discussed in Section \ref{sectionOver}.

(c) The cutoff frequency of the upper limit of the spinon continuum
is linear as a function of frustration for $J_2<0.35$. 

(d) The exponent of the cusp at the upper boundary of the spinon
continuum is always smaller than the value of  $\alpha=0.5$ predicted
for the two-spinon contribution for $J_2$. The value of $\alpha$
vanishes for $J_2\approx 0.2$ in agreement with the previous
observation\cite{YS97} that for that value the spectral
properties of the frustrated Heisenberg chain are similar to the
conformally invariant Haldane-Shastry\cite{Hald88,Shas88} model. 

The prefactor $\chi_0$ from Eq.\ (\ref{ChiCFT}) is of order 1 and
slightly frustration dependent. The values $(J_2,\chi_0)$ are: $(0,1.31(5))$,
$(0.15,1.12(5))$, $(0.2411,1.01(5))$, and $(0.35,0.88(5))$. For
$J_2=0.5$ one has $v^{1-2x}\chi_{\rbx{0}}=0.69(5)$. Values for
$J_2=0.45$ are not computed because of the peculiar finite size
effects shown in Figure \ref{ChilowT}. 

\begin{figure}[bt]
   \epsfxsize=0.48\textwidth
   \centerline{\epsffile{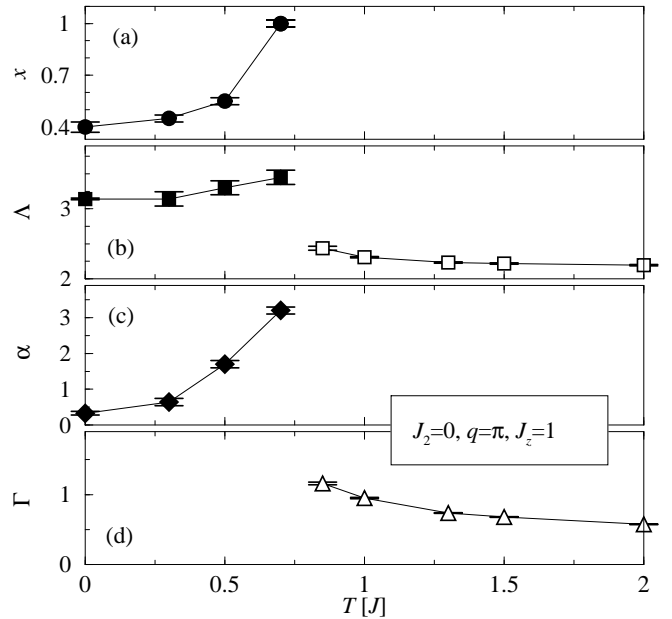}}
   \centerline{\parbox{\textwidth}{\caption{\label{ResultsTa000}
   \sl Extracted values for (a) the scaling dimension $x$ for $T<T^*$,
   (b) the upper continuum cutoff frequency for $T<T^*$ (full
   symbols) and the effective upper continuum edge in the
   diffusive/weakly-interacting-fermion regime (open symbols), (c) the
   cutoff exponent, and (d) the control parameter $\Gamma$ determining
   the decay as a function of time in the
   diffusive/weakly-interacting-fermion regime.}}}   
   \end{figure}

Figure \ref{ResultsTa000} summarizes the temperature dependence of
the fit parameters for the experimentally most relevant unfrustrated
chain with $J_2=0$. 

(a) The scaling variable approaches the value of the $XY$ model limit at
the crossover temperature to the diffusive regime $T^*\approx
0.7$. The direct determination of $x(T>T^*)$ is not possible but since
for $T\to\infty$ the weakly interacting fermion case is recovered it is
expected to lock in at $x(T>T^*)=1$.

(b) The upper continuum edge $\Lambda(T<T^*)$ marks a sharp cutoff
(full symbols) while in the diffusive regime (open symbols) it is the
effective, thermally smeared out upper continuum boundary. In the
weakly interacting fermion limit it saturates at
$\Lambda(T\to\infty)=2.15(1)$.

(c) The exponent of the upper continuum edge $\alpha$ increases with
increasing temperature reflecting the thermal smearing out of the
singularity. Its value at $T=T^*$ is so large that the cutoff is
barely singular and thus not very well defined. In the diffusive
regime ($T>T^*)$ this quantity is undefined. 

(d) $\Gamma$ is an effective parameter that controls the continuous
transition of the system from the diffusive behavior at $T^*$ to the
weakly interacting fermion limit at $T\to\infty$. For $J_2=0$ it
saturates at $\Gamma(T\to\infty) = 0.375(2)$

%%%%%%%%%%%%%%%%%%%%%%%%%%%%%%%%%%%%%%%%%%%%%%%%%%%%%%%%%%%%%%%%%%%%
%%%%%%%%%%%%%%%%%%%%%%%%%%%%%%%%%%%%%%%%%%%%%%%%%%%%%%%%%%%%%%%%%%%%

\section{Conclusions}

We have compared numerical results from the exact diagonalization of
finite systems with results from conformal field theory together
with implications from the density of states, the exactly solvable $XY$
model, and Bethe ansatz solutions for integrable systems. We use the
different finite size accuracy of the imaginary, real, and time
representations of the spin-correlation functions to extract reliable
information on the thermodynamic limit.

At low temperatures the dynamical correlation functions of
frustrated Heisenberg chains are well described by a multiplicative
superposition of the contribution from low lying elementary excitations
described by conformal field theory and a density of states and matrix
element induced singularity near the upper edge of the two-spinon
continuum. At the frustration value of $J_2\approx 0.2$ the system is
closest to the conformally invariant Haldane-Shastry model.

At $T^*\approx0.7$ we observe the crossover from the low temperature,
conformally invariant regime to a diffusive regime. All correlations
in time then decay exponentially. The diffusive regime connects
continuously to the weakly-interacting-fermion regime for
$T\to\infty$.

We give the frustration dependence of the control parameters for the
line shapes of the spin-correlation functions at $T=0$ and their
temperature dependence for the experimentally most relevant case of
$J_2=0$. The temperature dependence of the first moment sum rule of 
the spin-correlation function is accurately determined.

%%%%%%%%%%%%%%%%%%%%%%%%%%%%%%%%%%%%%%%%%%%%%%%%%%%%%%%%%%%%%%%%%%%%
%%%%%%%%%%%%%%%%%%%%%%%%%%%%%%%%%%%%%%%%%%%%%%%%%%%%%%%%%%%%%%%%%%%%

\section{Acknowledgments}

We thank I.A.\ Zaliznyak, J.\ Stolze, V.J.\ Emery, and M.\
Weinert for stimulating and instructive discussions. The work
performed at BNL was supported by DOE contract number
DE-AC02-98CH10886.

\bibliographystyle{prsty}
\bibliography{../../LIB/references}

\begin{thebibliography}{10}

\bibitem{KM97}
M. Karbach and G. M{\"u}ller, Comp. in Phys. {\bf 11},  36  (1997).

\bibitem{KHM98}
M. Karbach, K. Hu, and G. M{\"u}ller, Comp. in Phys. {\bf 12},  565  (1998).

\bibitem{SSS97b}
O.~A. Starykh, A.~W. Sandvik, and R.~R.~P. Singh, Phys. Rev. B {\bf 55},  14953
   (1997).

\bibitem{FLS97}
K. Fabricius, U. L{\"o}w, and J. Stolze, Phys. Rev. B {\bf 55},  5833  (1997).

\bibitem{YS97}
H. Yokoyama and Y. Saiga, J. Phys. Soc. Jpn. {\bf 66},  3617  (1997).

\bibitem{FL98}
K. Fabricius and U. L{\"o}w, Phys. Rev. B {\bf 57},  13371  (1998).

\bibitem{MTBB81}
G. M{\"u}ller, H. Thomas, H. Beck, and J.~C. Bonner, Phys. Rev. B {\bf 24},
  1429  (1981).

\bibitem{Schu86}
H.~J. Schulz, Phys. Rev. B {\bf 34},  6372  (1986).

\bibitem{Tsve95}
A.~M. Tsvelik, {\em Quantum field theory in condensed matter physics}
  (Cambridge University Press, Cambridge, 1995).

\bibitem{EN88}
V.~J. Emery and C. Noguera, Phys. Rev. Lett. {\bf 60},  631  (1988).

\bibitem{CPK+95}
R. Chitra {\it et~al.}, Phys. Rev. B {\bf 52},  6581  (1995).

\bibitem{YMV96}
Y. Yu, G. M{\"u}ller, and V. Viswanath, Phys. Rev. B {\bf 54},  9242  (1996).

\bibitem{KMB+97}
M. Karbach {\it et~al.}, Phys. Rev. B {\bf 55},  12510  (1997).

\bibitem{SSS97a}
O.~A. Starykh, R.~R.~P. Singh, and A.~W. Sandvik, Phys. Rev. Lett. {\bf 78},
  539  (1997).

\bibitem{KHM00}
M. Karbach, K. Hu, and G. M{\"u}ller, cond-mat/0008018 (unpublished)  (2000).

\bibitem{KM00}
M. Karbach and G. M{\"u}ller, Phys. Rev. B {\bf 62},  14871  (2000).

\bibitem{Wern01a}
R. Werner, Phys. Rev. B {\bf 63},  in print  (2001).

\bibitem{ACH+95}
T. Ami {\it et~al.}, Phys. Rev. B {\bf 51},  5994  (1995).

\bibitem{MEU96}
N. Motoyama, H. Eisaki, and S. Uchida, Phys. Rev. Lett. {\bf 76},  3212
  (1996).

\bibitem{CTC+96}
R. Coldea {\it et~al.}, J. Phys.: Condens. Matter {\bf 8},  7473  (1996).

\bibitem{TCNT95}
D.~A. Tennant, R.~A. Cowley, S.~E. Nagler, and A.~M. Tsvelik, Phys. Rev. B {\bf
  52},  13368  (1995).

\bibitem{Wern99}
R. Werner, {\em The spin-Peierls transition in $\mathrm{CuGeO}_3$} (Ph.D.
  thesis, Dortmund, 1999),
  http://eldorado.uni-dortmund.de:8080/FB2/ls8/forschung/1999/werner.

\bibitem{FG97}
A. Fledderjohann and C. Gros, Europhys. Lett. {\bf 37},  189  (1997).

\bibitem{Frad91}
E.~F. Fradkin, {\em Field Theories of Condensed Matter Systems}
  (Addison-Wesley, New York, 1991).

\bibitem{Eme79}
V.~J. Emery,  in {\em Highly conducting one-dimensional solids}, edited by
  J.~T. Devreese, R.~P. Evrard, and V.~E. van Doren (Plemun, New York, 1979),
  p.\ 247.

\bibitem{NO94}
K. Nomura and K. Okamoto, J. Phys. A: Math. Gen. {\bf 27},  5773  (1994).

\bibitem{boundspecquote}
The requirement of the spectrum to be bounded can be relaxed to
  $\lim_{\omega\to\infty} \Im\chi(q,\omega) \sim \omega^{-s}$ with $s>2$. For
  $s=2$ there will be logarithmic corrections $\lim_{\omega\to\infty}
  \Re\chi(q,\omega) \sim \omega^{-2} \ln[\omega^{-1}]$.

\bibitem{MG69}
C.~K. Majumdar and D.~K. Ghosh, J. Math. Phys. {\bf 10},  1388  (1969).

\bibitem{degenquote}
Interestingly, the spectral lines from the singlet excitations relevant for the
  dimer-dimer correlation funtions\protect\cite{Wern01a} and the triplet
  excitations are degenerate for $J_2=J_c$.

\bibitem{lowomegaquote}
For practical purposes this discrepancy can be overcome for small frequencies
  by rescaling the frequency dependence as has been done in the case of
  dimer-dimer correlation functions (see Ref.\ \onlinecite{Wern01a}).

\bibitem{dCP62}
J. des Cloizeaux and J.~J. Pearson, Phys. Rev. {\bf 128},  2131  (1962).

\bibitem{SS81}
B.~S. Shastry and B. Sutherland, Phys. Rev. Lett. {\bf 47},  964  (1981).

\bibitem{Uhri99}
G. Uhrig, {\em Niedrigdimensionale Spinsysteme und Spin-Phonon-Kopplung}
  (Habilitation, Cologne, 1999).

\bibitem{Vill74}
J. Villain, J. Phys. France {\bf 35},  27  (1974).

\bibitem{Hald88}
F.~D.~M. Haldane, Phys. Rev. Lett. {\bf 60},  635  (1988).

\bibitem{Shas88}
B.~S. Shastry, J. Stat. Phys. {\bf 50},  57  (1988).

\end{thebibliography}

\end{document}